\documentclass[twocolumn,prl,amsmath,amssymb, aps,superscriptaddress, nofootinbib]{revtex4-2}
 \usepackage{comment}
\usepackage[titletoc,toc,title]{appendix}
\usepackage{titletoc}
\usepackage{titlesec}
\usepackage[version=4]{mhchem} 
\usepackage[normalem]{ulem}
\usepackage{mhchem}
\usepackage{amsmath}
\usepackage{amssymb}
\usepackage{amsfonts}
\usepackage{graphicx}
\usepackage{xfrac}
\usepackage{footmisc}
\usepackage{comment}
\usepackage{pifont}
\usepackage{times}
\usepackage{enumitem}
\usepackage{centernot}
\usepackage{cancel}
\usepackage{slashed}
\usepackage{relsize}
\usepackage{amsmath,amssymb,mathrsfs}
\usepackage{times}
\usepackage{epsfig}
\usepackage{verbatim}
\usepackage{bm}
\usepackage[utf8]{inputenc}
\usepackage{graphics}
\usepackage{graphicx,epsfig,amssymb,amsmath,color,cancel}
\usepackage{subfigure}
\usepackage[english]{babel}
\usepackage{adjustbox}
\usepackage{physics}
\usepackage{mathtools, stmaryrd}
\usepackage{xparse} 
\usepackage{placeins}
\DeclarePairedDelimiterX{\Iintv}[1]{\llbracket}{\rrbracket}{\iintvargs{#1}}
\NewDocumentCommand{\iintvargs}{>{\SplitArgument{1}{,}}m}
{\iintvargsaux#1} %
\NewDocumentCommand{\iintvargsaux}{mm} {#1\mkern1.5mu..\mkern1.5mu#2}

\usepackage{array}
\newcolumntype{P}[1]{>{\centering\arraybackslash}p{#1}}
\newcolumntype{M}[1]{>{\centering\arraybackslash}m{#1}}
\usepackage{tabularray}

\usepackage[usenames,dvipsnames,svgnames,table]{xcolor}
\definecolor{blue_col}{RGB}{0,92,175}
\definecolor{red_col}{RGB}{203,64,66}

\usepackage[linktoc=page,bookmarks=false,colorlinks=true,linkbordercolor=PineGreen,citebordercolor=PineGreen,urlbordercolor=PineGreen]{hyperref}
\hypersetup{
    colorlinks=true,
    linkcolor=red_col,
    filecolor=Mahogany,      
    urlcolor=blue_col,
    citecolor=blue_col,
}

\usepackage{tabularray}
\UseTblrLibrary{booktabs}

\definecolor{zima_blue}{HTML}{1393C1}
\hypersetup{setpagesize=false,bookmarksnumbered=true,bookmarksopen=true,%
colorlinks=true,linkcolor=zima_blue,urlcolor=zima_blue,citecolor=zima_blue,linktocpage=false}

\usepackage{orcidlink}

\usepackage[
    starfontserif 
    ]{starfont}
\usepackage{amsmath}

\DeclareSymbolFont{starfontsym}{OT1}{sts}{m}{n}
\DeclareMathSymbol{\mathSun}{\mathord}{starfontsym}{115}
\DeclareMathSymbol{\mathMercury}{\mathord}{starfontsym}{102}
\DeclareMathSymbol{\mathVenus}{\mathord}{starfontsym}{103}
\DeclareMathSymbol{\mathTerra}{\mathord}{starfontsym}{76}
\DeclareMathSymbol{\mathvarTerra}{\mathord}{starfontsym}{108}
\DeclareMathSymbol{\mathMoon}{\mathord}{starfontsym}{100}
\DeclareMathSymbol{\mathvarMoon}{\mathord}{starfontsym}{97}
\DeclareMathSymbol{\mathMars}{\mathord}{starfontsym}{104}
\DeclareMathSymbol{\mathJupiter}{\mathord}{starfontsym}{106}
\DeclareMathSymbol{\mathSaturn}{\mathord}{starfontsym}{83}
\DeclareMathSymbol{\mathUranus}{\mathord}{starfontsym}{70}
\DeclareMathSymbol{\mathvarUranus}{\mathord}{starfontsym}{65}
\DeclareMathSymbol{\mathNeptune}{\mathord}{starfontsym}{71}
\DeclareMathSymbol{\mathPluto}{\mathord}{starfontsym}{74}
\DeclareMathSymbol{\mathvarPluto}{\mathord}{starfontsym}{72}

\titleformat{\section}
{\normalfont\fontsize{10}{12}\bfseries  \centering }{\thesection.}{1em}{}
\titleformat{\subsection}
{\normalfont\fontsize{10}{12}\bfseries \centering}{\thesubsection.}{1em}{}
\titleformat{\subsubsection}
{\normalfont\fontsize{10}{12}\bfseries \centering}{\thesubsubsection)}{1em}{}

\titleformat{\paragraph}
{\normalfont\fontsize{10}{12}\bfseries  }{\thesection:}{1em}{}
\clearpage

\begin{document}

\preprint{}

\title{Primordial Black Holes and Wormholes from Domain Wall Networks}

\author{Yann Gouttenoire~\orcidlink{0000-0003-2225-6704}}
\email{yann.gouttenoire@gmail.com}
\affiliation{School of Physics and Astronomy, Tel-Aviv University, Tel-Aviv 69978, Israel}
\author{Edoardo Vitagliano~\orcidlink{0000-0001-7847-1281}}
\email{edoardo.vitagliano@mail.huji.ac.il}
\affiliation{Racah Institute of Physics, Hebrew University of Jerusalem, Jerusalem 91904, Israel}

\begin{abstract}

Domain walls (DWs) are topological defects originating from phase transitions in the early universe. In the presence of an energy imbalance between distinct vacua, enclosed DW cavities shrink until the entire network disappears. 
By studying the dynamics of thin-shell bubbles in General
Relativity, we demonstrate that closed DWs with sizes exceeding the cosmic horizon tend to annihilate later than the average. This delayed annihilation allows for the formation of large overdensities, which, upon entering the Hubble horizon, eventually collapse to form Primordial Black Holes (PBHs). We rely on 3D percolation theory to calculate the number density of these late-annihilating DWs, enabling us to infer the abundance of PBHs. A key insight from our study is that DW networks with the potential to emit observable Gravitational Waves are also likely to yield detectable PBHs. Additionally, we find that wormholes connected to baby-universes can be produced and conclude on the possibility to generate a multiverse.

\end{abstract}
\maketitle
\raggedbottom

\section{INTRODUCTION}
\label{sec:intro}
The study of primordial black holes (PBHs) has become a vibrant field of exploration since the 2015 discovery of gravitational waves from solar-mass black holes mergers~\cite{LIGOScientific:2018mvr}. Any observation of black holes with masses below that of the Sun would be a smoking-gun for the gravitational collapse of large density fluctuation, present in the initial cosmic plasma~\cite{Carr:1974nx}. 
An intriguing possibility involves the inhomogeneities generated by domain walls (DWs) either under the form of networks~\cite{Vachaspati:2017hjw,Ferrer:2018uiu,Gelmini:2022nim,Gelmini:2023ngs} or nucleated during inflation \cite{Deng:2016vzb,Deng:2017uwc,Deng:2020mds,Liu:2019lul,Kusenko:2020pcg,Ge:2023rrq}. 

DW networks are topological defects that form during cosmological phase transitions in the early universe when nearly-degenerate vacua are present. Shortly after its formation, the DW network evolves in a scaling regime during which the correlation length approximately equals the horizon size $L\sim t$ \cite{Press:1989yh}, and the fraction of the total universe energy density stored in DWs increases linearly with time $\rho_{\rm DW}/\rho_{\rm tot} \propto t$~\cite{Press:1989yh}. A universe dominated by DWs would lead to a universe either primarily filled with black holes or to eternally-inflating universes, depending on whether the observer location lies within a true vacuum or false vacuum region~\cite{Vilenkin:2000jqa}.
A graceful exit from this daunting scenario could occur if there is an energy bias $V_{\rm bias}$ between the different vacua, as we review in Sec.~\ref{sec:biased_net}. After a time $t_{\rm ann}$, the vacuum energy difference $V_{\rm bias}$ counterbalances the pressure due to the wall tension $\sigma$, driving DWs towards annihilation before they can dominate the universe at a time $t_{\rm dom}$~\cite{Kibble:1976sj,Vilenkin:1981zs,Sikivie:1982qv,Gelmini:1988sf}. During this annihilation phase, closed DWs  shrink and under specific conditions, could enter within their Schwarzschild radius and form PBHs~\cite{Ferrer:2018uiu,Deng:2016vzb,Deng:2017uwc,Deng:2020mds}, a process  dubbed ``catastrogenesis'' in~\cite{Gelmini:2022nim,Gelmini:2023ngs}. As explained above, if $t_{\rm ann}\gtrsim t_{\rm dom}$, the universe is filled with PBHs and wormholes connected to baby-universes \cite{Vilenkin:2000jqa}. Per continuity, we conclude that there must exist a region in the ballpark $t_{\rm ann}\lesssim t_{\rm dom}$, where DWs networks produce just enough PBHs to explain dark matter (DM), or just enough wormholes to generate a multiverse in our past lightcone \cite{Garriga:2015fdk,Garriga:1997ef,Linde:2015edk}. 

\begin{figure}[h!]
\centering
\includegraphics[width=1\linewidth]{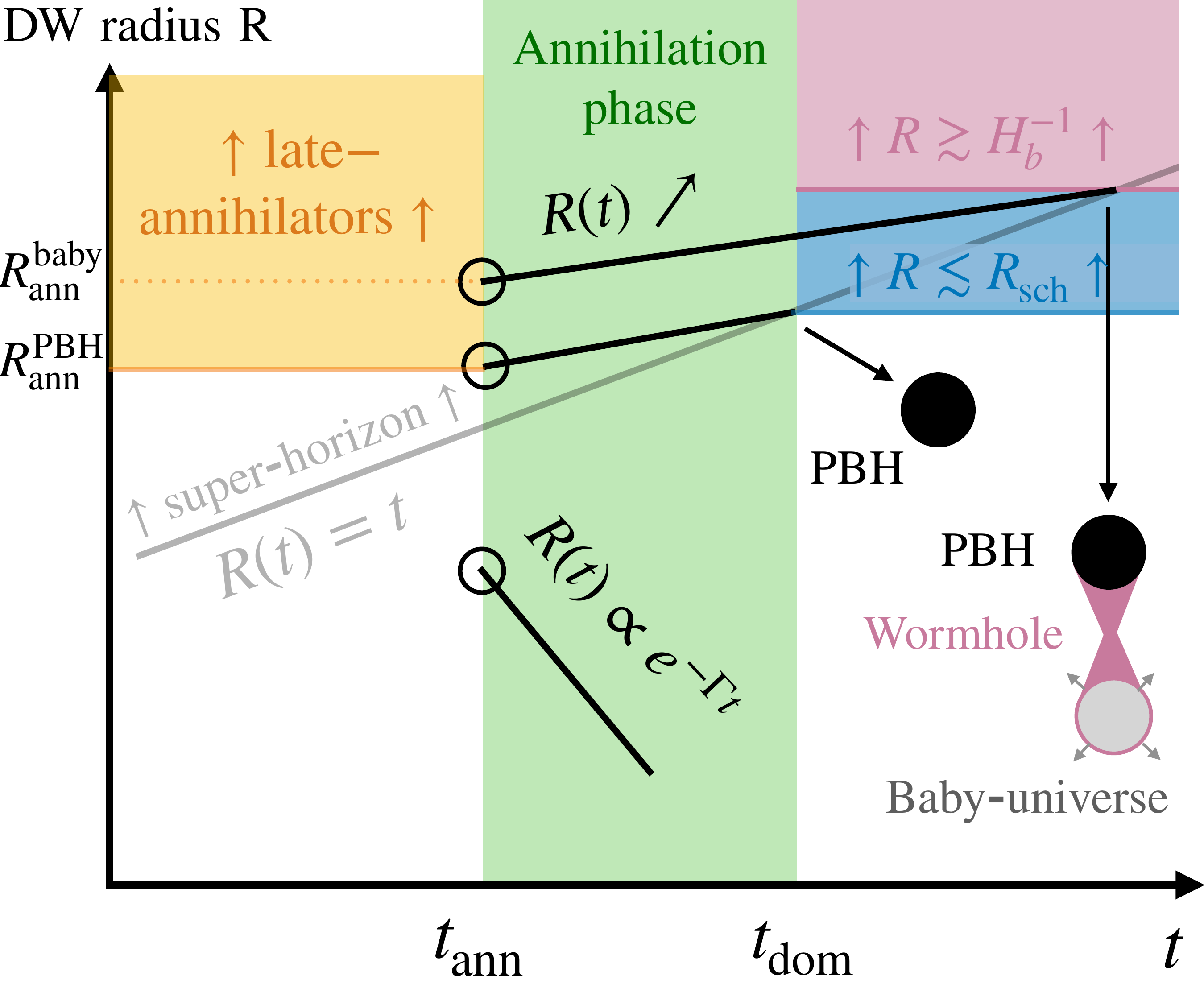}
\caption{\small   \label{fig:PBH_from_DW_drawing} \textbf{{\normalsize Domain wall late-annihilation mechanism.}} During annihilation (\textbf{Green} region), closed DWs with sub-horizon size shrink exponentially fast while super-horizon DWs continue to grow due to Hubble flow. Closed DW configurations with a radius larger than $R_{\rm ann}^{\rm PBH}$ at the onset of the annihilation phase, which are dubbed ``late-annihilators'' in the \textbf{Orange} region, enter inside their Schwarzschild radius $R\lesssim 2GM$ in the \textbf{Blue} region, leading them to collapse into PBHs. DWs with radius larger than $R_{\rm ann}^{\rm baby}$ become larger than the cosmological horizon $H_b^{-1}$ of an inflating universe with same vacuum energy $V_{\rm bias}$ as their interior, in the  \textbf{Purple} region, with $H_b^2 = V_{\rm bias}/3M_{\rm Pl}^2$. They become eternally-inflating baby-universes detaching from the parent universe through wormholes and leaving PBHs behind. }
\end{figure}
Correcting for mistakes of previous work \cite{Ferrer:2018uiu,Gelmini:2022nim,Gelmini:2023ngs}, we calculate the abundance of PBHs, and for the first time the number of those which hid behind their horizon a wormhole connected to an eternally-inflating baby-universe. Previous studies~\cite{Ferrer:2018uiu,Gelmini:2022nim,Gelmini:2023ngs} incorrectly assumed that the radius of DWs grows as $R\simeq t$. Using energy arguments, Ref.~\cite{Gelmini:2023kvo} commented that the relation $R\simeq t$ is not applicable during the annihilation phase. Our results suggest that $R\simeq t$ is never valid, neither during the annihilation phase nor at any other period.  In Sec.~\ref{sec:DW_EoM}, we solve the equation of motion (EoM) of a single spherical DW in an expanding universe, derived from Einstein equations in the thin-shell limit. 
The main results are sketched in Fig.~\ref{fig:PBH_from_DW_drawing}.
 The radius grows as $R\propto a$ or slower if superhorizon $R\gtrsim t$, but quickly shrink if sub-horizon $R\lesssim t$.  We find that PBHs enter their Schwarzschild radius if they have a radius larger than $R\gtrsim R_{\rm ann}^{\rm PBH}>t$ just before the annihilation phase.  While most of DWs shrink and decay away into scalar waves, those DWs with super-horizon size, which we call \textit{late-annihilators}, continue to grow during the annihilation phase until they finally collapse when entering inside the cosmic horizon. We also derive a critical radius $R_{\rm ann}^{\rm baby}$ larger than $R_{\rm ann}^{\rm PBH}$ by about $30\%$ above which DWs collapse into wormholes connected to eternally-inflating baby-universes.
We find that the PBH formation threshold $R_{\rm ann}^{\rm PBH}$ is highly sensitive to the surface energy of DWs, which can have Lorentz factors up to $\gamma\sim 10$, and to gravitational binding energies. These contributions, which constitute $75\%$ of the PBH mass, were overlooked in prior studies \cite{Ferrer:2018uiu,Gelmini:2022nim,Gelmini:2023ngs}. In Sec.~\ref{app:percolation_theory}, we employ for the first time percolation theory to estimate the fraction of closed DWs which are large enough to host a false vacuum ball of radius $R_{\rm min}$. In Sec.~\ref{sec:PBH_WH_abundance}, we deduce the abundance of PBHs and wormholes. In Sec.~\ref{sec:GW}, we review the Stochastic Gravitational Wave Background (SGWB) produced by annihilating DW networks. We point the complementarity between GW detectability by future observatories, PBHs, and baby-universe production.

\section{BIASED NETWORK EVOLUTION}
\label{sec:biased_net}
Initially, the work of the DW surface tension $\sigma$, with equivalent pressure $\mathcal{P}_T=C_d\, \sigma / L $ toward straightening, is dampened by the friction pressure $\mathcal{P}_V \simeq \,\beta\, T^4$. The quantity $L$ is the correlation length of the DW network, $C_d$ and $\beta$ are dimensionless parameters setting respectively the strength of the straightening pressure and of the DW-plasma interactions. Estimating DWs to have typical curvature radius $L \simeq v t$, one obtains that DWs start moving with relativistic velocity $v\simeq \mathcal{O}(0.1)$ after the time
\begin{equation}
t_{\rm rel} \simeq 2 \times 10^{-3}\left(\frac{v}{0.1} \right)\left(\frac{100}{g_\star} \right)\frac{\beta}{C_d }\frac{M_{\rm Pl}^2}{\sigma}.
\end{equation}
where $M_{\rm Pl}=1/\sqrt{8\pi G}\simeq2.435 \times 10^{18}\,\rm GeV$ is the reduced Planck mass, $g_\star$ the number of relativistic degrees of freedom and where we used Friedmann's equation $T=1.2\sqrt{M_{\rm Pl}/t}/g_\star^{1/4}$.
The magnitude of the friction coefficient, denoted by $\beta$, varies depending on the particle physics model  \cite{Blasi:2022ayo,Blasi:2023sej}. In our study, we assume $\beta \ll 1$ and briefly touch upon the implications of friction on PBH formation towards the end of the paper in Sec.~\ref{sec:further_consideration}.
We assume that the energy bias $V_{\text{bias}}$ is initially insignificant at the time of DW formation and remains negligible when they begin to move without friction. Under these conditions, numerical simulations have shown that the energy density of DWs quickly reaches the scaling regime,
\begin{equation}
\label{eq:rho_DW}
    \rho_{\rm DW} = \frac{\sigma}{L},\qquad L=  t/\mathcal{A},
\end{equation} 
where $ \mathcal{A}$ is the area parameter.  Based on numerical simulations, the area parameter for a $\mathcal{Z}_2$ symmetric model is approximately $ \mathcal{A} \simeq 0.8 \pm 0.1$~\cite{Hiramatsu:2013qaa}. In contrast, $\mathcal{Z}_{N_{\rm DW}}$ symmetric models with $N_{\rm DW}>2$, where cosmic strings attached to $N_{\rm DW}$ DWs are present, tend to yield larger values $\mathcal{A} = (0.37\pm 0.04)N_{\rm DW}$~\cite{Kawasaki:2014sqa}.
 DWs annihilate when the vacuum pressure $\mathcal{P}_V=V_{\rm bias}$ dominates over the pressure $\mathcal{P}_T=C_d\, \sigma / L $ arising from their surface tension $\sigma$, after the time
\begin{equation}
\label{eq:t_ann}
t_{\rm ann}~ \simeq~  C_d \,\mathcal{A}\,\frac{\sigma}{V_{\rm bias}},
\end{equation}
where the factor $C_d\simeq \mathcal{O}(\rm a~few)$ can be inferred from numerical simulations~\cite{Kawasaki:2014sqa} and which we set to $C_d\simeq 3$. To prevent entering the catastrophic scenario of DW domination, DW annihilation must proceed before the network dominates the energy budget of the universe, occurring when $3M_{\rm Pl}^2H^2\simeq V_{\rm bias}$ around the time
\begin{equation}
\label{eq:t_dom}
    t_{\rm dom }~ \simeq~ \frac{\sqrt{3}M_{\rm Pl}}{2\sqrt{V_{\rm bias}}}~=~1.5\times 10^{-11}{~\rm s} \left(\frac{300~\rm GeV}{V^{1/4}_{\rm bias}} \right)^{\!2},
\end{equation}
which upon comparing to Eq.~\eqref{eq:t_ann}, leads to the condition
\begin{equation}
    V_{\rm bias} ~\gtrsim ~\frac{4}{3}\left(\hspace{-0.1cm}\mathcal{A\,}C_d \frac{\sigma}{M_{\rm Pl}}\right)^{\!2} \hspace{-0.1cm}\simeq (300~\rm GeV)^4\mathcal{A\,}^{\!2}C_d^2 \left(\frac{\sigma^{1/3}}{58~\rm PeV}\right)^{\!6}\hspace{-0.1cm}.
\end{equation}
In the absence of a bias, the DW network would dominate when $3M_{\rm Pl}^2H^2\simeq \sigma/L$ after the time
\begin{equation}
\label{eq:t_dom_no_bias}
    t_{\rm dom }^{\mathsmaller{\rm unbias}}~ =~ \frac{3M_{\rm Pl}^2}{4\mathcal{A} \sigma}~=~\frac{1.5\times 10^{-11}{~\rm s}}{\mathcal{A}} \left(\frac{58~\rm PeV}{\sigma^{1/3}} \right)^{\!3}.
\end{equation}

\begin{figure*}[ht!]
\centering
\begin{adjustbox}{max width=1\linewidth,center}
\raisebox{0cm}{\makebox{\includegraphics[ width=0.49\textwidth, scale=1]{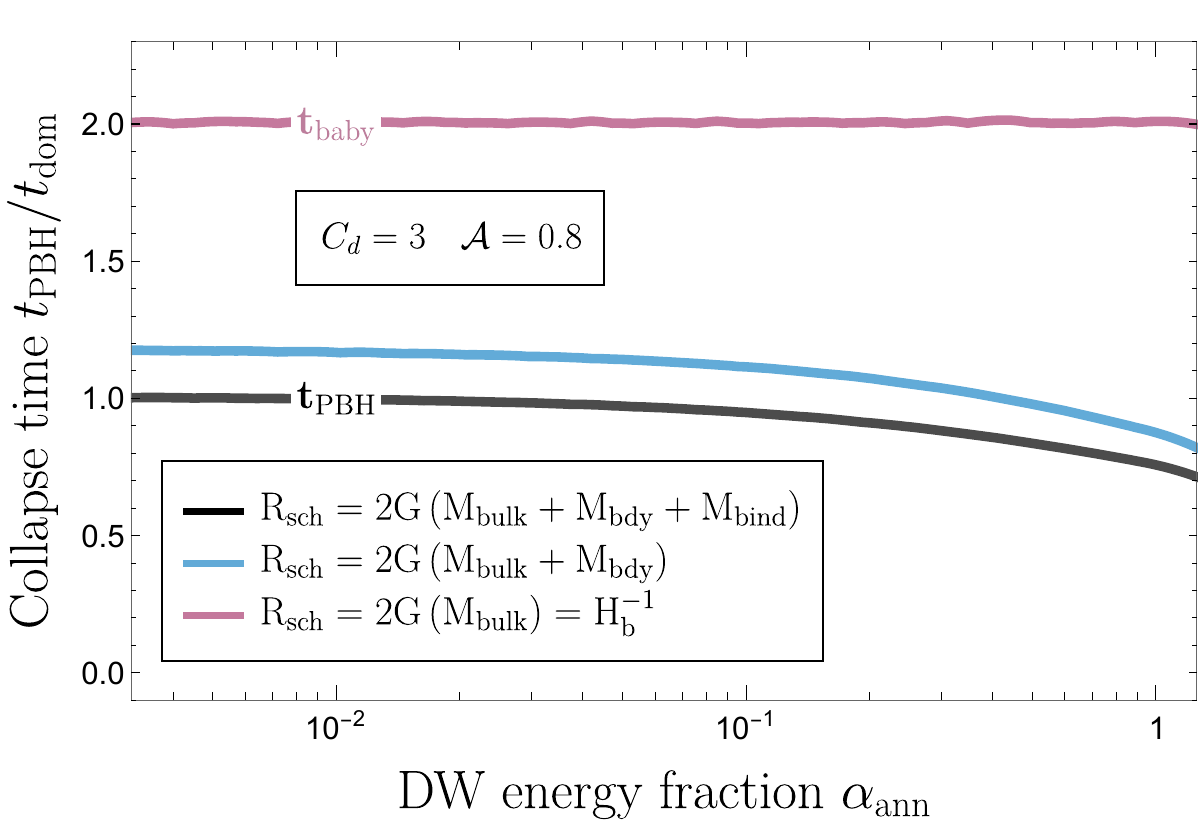}}}
\raisebox{0cm}{\makebox{\includegraphics[ width=0.49\textwidth, scale=1]{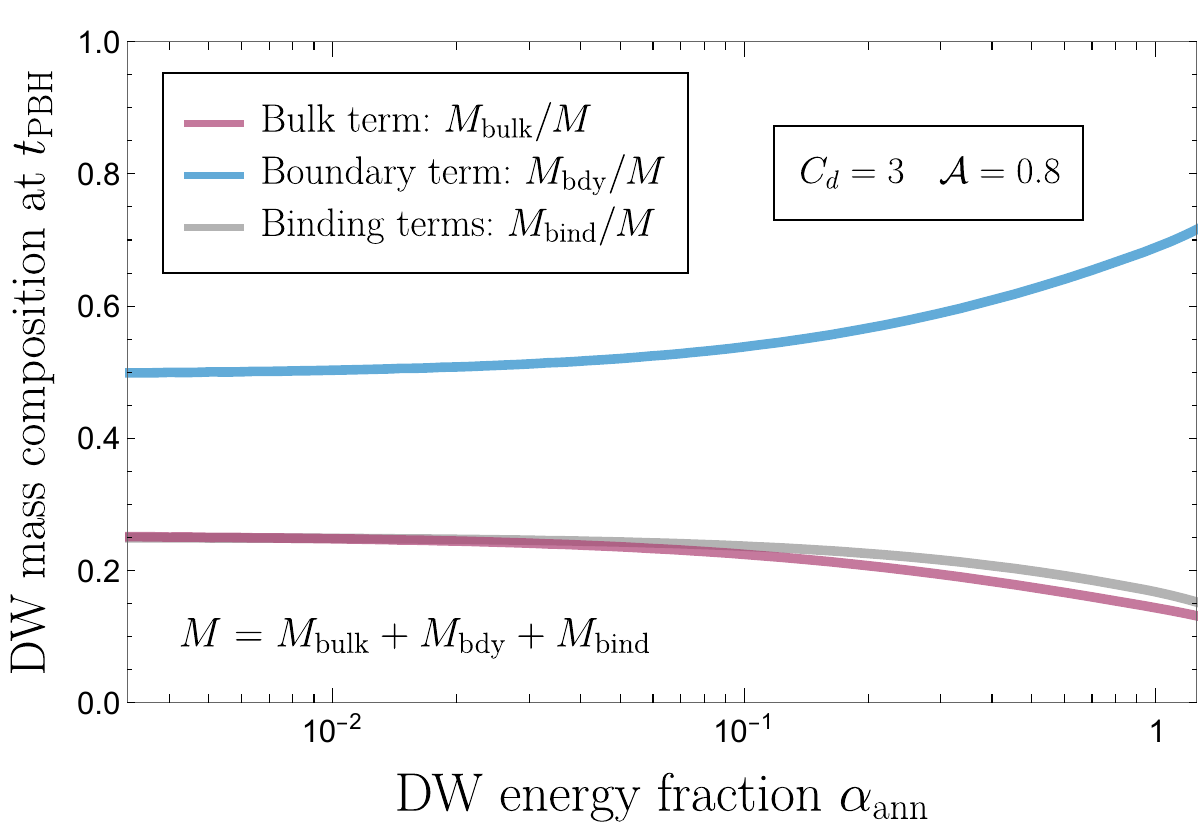}}}
\end{adjustbox}
\caption{ \small  \label{fig:collapse_time} {LEFT:}  We show the time $t_{\rm PBH}$ at which a spherical DW collapses into PBH, found from integrating the equations of motion in Eqs.~\eqref{eq:Friedmann_0} and \eqref{eq:EoM} with the boundary conditions in Eq.~\eqref{eq:IC} and solving Eq.~\eqref{eq:ptRadius}, for different levels of approximation of the DW mass, either accounting for volume (``bulk''), surface (``bdy'') and gravitational binding (``bind'') terms in \textbf{black}, or neglecting binding term in \textbf{blue} or neglecting both binding and surface terms in \textbf{purple}. The purple line, where only the vacuum energy is included, also corresponds to the time $t_{\rm baby}$ in Eq.~\eqref{eq:t_baby} of formation of a baby-universe. {RIGHT:} Composition of the DW mass at the time $t_{\rm PBH}$ when it enters its Schwarzschild radius. The bulk component only accounts for $25~\%$ of the DW mass. Another $25\%$ is given by gravitational binding energies $M_{\rm bind}$ and $50\%$ of the DW mass is given by its surface term $M_{\rm bdy}$. }
\end{figure*}
\section{TIME EVOLUTION OF A SPHERICAL DW}
\label{sec:DW_EoM}

\subsection{Equation of motion}
The DW network contains closed configurations of size $R(t)$ which we treat as spherical for simplicity. We discuss the implication of such approximation later in Sec.~\ref{sec:further_consideration}.
We denote by $R(t)=a(t)\chi(t)$ the physical radius of a spherical DW and by $\chi(t)$ its comoving radius. We introduce the wall Lorentz factor $\gamma=1/\sqrt{1-(a\dot{\chi})^2}$. The EoM of a spherical DW can be determined through the Israel junction conditions~\cite{Berezin:1982ur,Berezin:1987bc,Blau:1986cw,Maeda:1985ye,Tanahashi:2014sma,Deng:2016vzb,Deng:2017uwc,Deng:2020mds}, named after Werner Israel~\cite{1966NCimB..44....1I}. These conditions equate the discontinuity in the extrinsic curvature across the wall to the wall surface energy-momentum tensor~\cite{Israel:1966rt}. For a shrinking DW, one obtains~\cite{Deng:2020mds}
\begin{multline}
\label{eq:EoM_0}
    \ddot{\chi}+(4-3a^2\dot{\chi}^2)H\dot{\chi} + \frac{2}{a^2\chi}(1-a^2\dot{\chi}^2)\\
    =-\left(\frac{V_{\rm bias}}{\sigma}-6\pi G\sigma\right)\frac{(1-a^2\dot{\chi}^2)^{3/2}}{a},
\end{multline}
where $a(t)$ is the scale factor of the universe inside the DW,
\begin{equation}
\label{eq:Friedmann_0}
    \dot{a}(t)=H(t)a(t),\qquad  H^2(t) = \frac{\rho_{\rm rad}(t)+V_{\rm bias}}{3M_{\rm Pl}^2},
\end{equation}
with $\rho_{\rm rad}(t) = \rho_{\rm rad}(t_{\rm ann})\left[a(t_{\rm ann})/a(t)\right]^4$.
Introducing the dimensionless quantities
\begin{equation}
    \tilde{\chi}(\tau) \equiv \chi(t/t_{\rm ann})/t_{\rm ann},\qquad \tau \equiv t/t_{\rm ann},
\end{equation}
the EoM in Eq.~\eqref{eq:EoM_0} can be rewritten as
\begin{multline}
\label{eq:EoM}
    \ddot{\tilde{\chi}}+(4-3a^2\dot{\tilde{\chi}}^2)\tilde{H}\dot{\tilde{\chi}} + \frac{2}{a^2\tilde{\chi}}(1-a^2\dot{\tilde{\chi}}^2)\\
    =-\left(\mathcal{A}C_d-\frac{9\alpha_{\rm ann}}{16\mathcal{A}}\right)\frac{(1-a^2\dot{\tilde{\chi}}^2)^{3/2}}{a},
\end{multline}
where the dot now denotes the derivative with respect to $\tau$, $\tilde{H}$ is defined below, and $\alpha_{\rm ann}$ is the DW network energy fraction at $t_{\rm ann}$
\begin{equation}
\label{eq:alpha_ann}
    \alpha_{\rm ann} \equiv \frac{\rho_{\rm DW}}{\rho_{\rm tot}}\Big|_{t=t_{\rm ann}}\simeq \left( \frac{t_{\rm ann}}{t_{\rm dom }^{\mathsmaller{\rm unbias}}} \right) \simeq C_d^{-1} \left(\frac{t_{\rm ann}}{ t_{\rm dom}}\right)^2,
\end{equation}
where we used $\rho_{\rm tot}\simeq 3M_{\rm Pl}^2/4t^2$ and Eqs.~(\ref{eq:rho_DW}, \ref{eq:t_ann}, \ref{eq:t_dom}, \ref{eq:t_dom_no_bias}). Upon approximating $\rho_{\rm rad}(t_{\rm ann})\simeq \rho_{\rm tot}(t_{\rm ann})$, which is exact if $t_{\rm dom}\gg t_{\rm ann}$, the dimensionless Hubble factor reads
\begin{equation}
    \tilde{H}(\tau)=\sqrt{\left(a(t_{\rm ann})/a(t)\right)^4+C_d \alpha_{\rm ann}}/2.
\end{equation}
We determine the evolution of closed DWs assuming that their comoving motion is initially frozen with a radius $R_i$ much larger than the horizon $R_i\gg t_i$ much before the onset of annihilation $t_i\ll t_{\rm ann}$. We solve the equation of motion in Eq.~\eqref{eq:EoM} and Friedmann's equation in Eq.~\eqref{eq:Friedmann_0} with initial conditions
\begin{equation}
\label{eq:IC}
    a(t_i)\chi(t_i)=R_i,\quad \chi'(t_i)=0,\quad a(t_i) = 1,
\end{equation}
with $R_i \gg t_i$ and $t_i \ll t_{\rm ann}$. The initial radius $R_i$ labels the different trajectories. Examples of DW trajectories are shown in Fig.~\ref{fig:Radius_EoM}-left.
\subsection{PBH collapse time}
\begin{figure}[h!]
\centering
\begin{adjustbox}{max width=1\linewidth,center}
\raisebox{0cm}{\makebox{\includegraphics[ width=0.49\textwidth, scale=1]{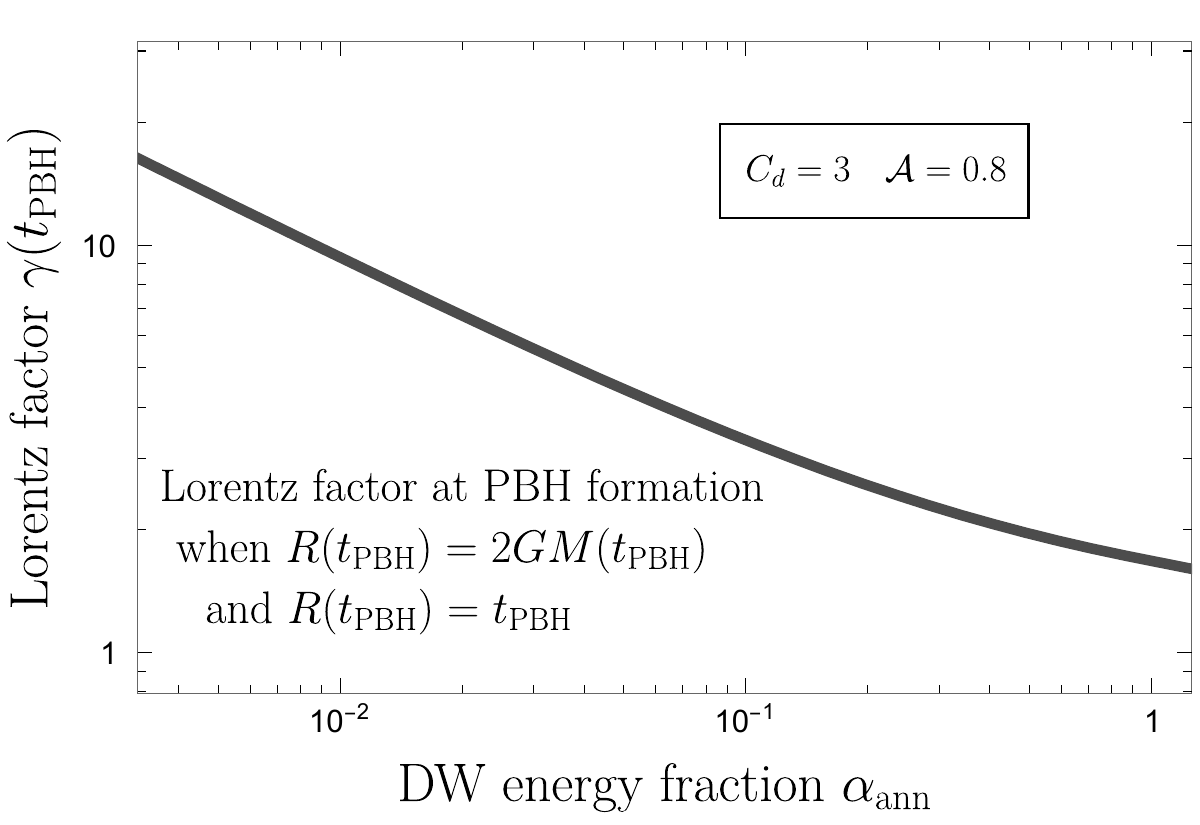}}}
\end{adjustbox}
\caption{ \small  \label{fig:Lorentz_factor}Lorentz factor $\gamma =1/\sqrt{1-(a\dot{\chi})^2}$ of shrinking DW at the time $t_{\rm PBH}$ of collapse into PBH.}
\end{figure}

\begin{figure*}[ht!]
\centering
\begin{adjustbox}{max width=1\linewidth,center}
\raisebox{0cm}{\makebox{\includegraphics[ width=0.49\textwidth, scale=1]{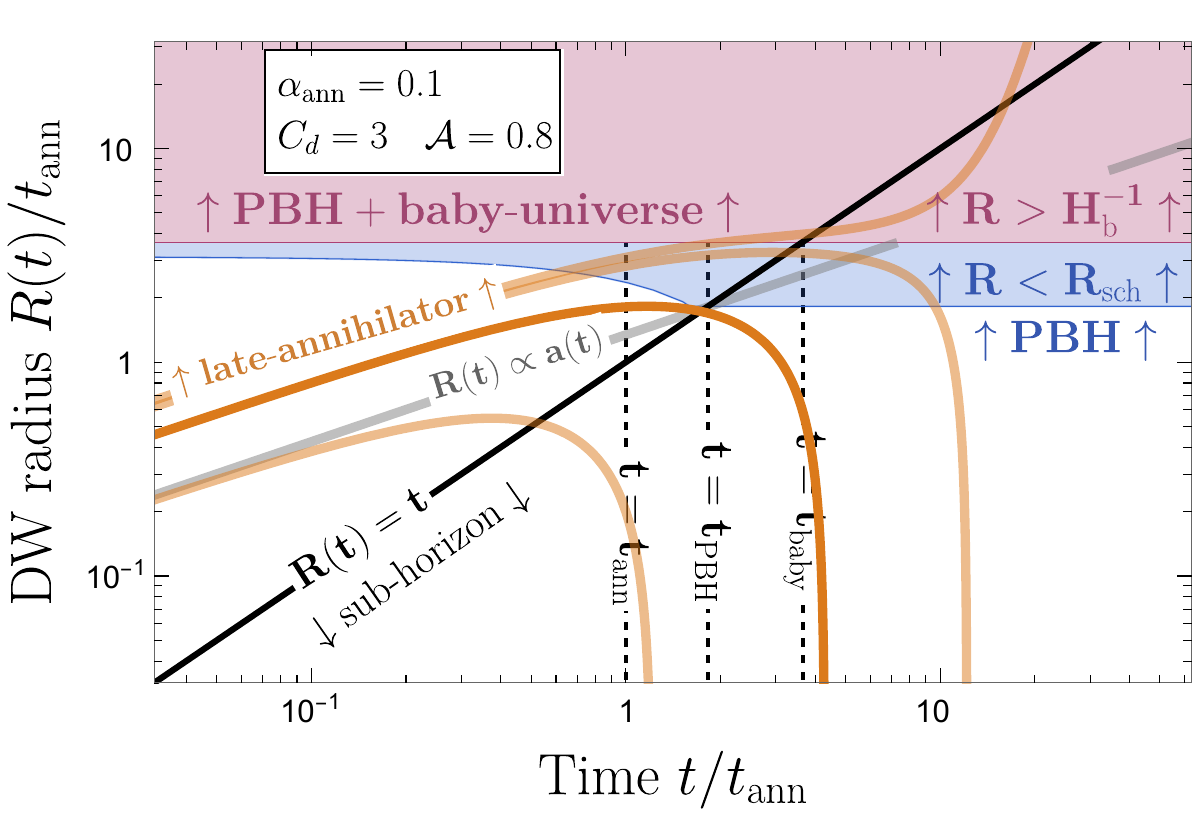}}}
\raisebox{0cm}{\makebox{\includegraphics[ width=0.49\textwidth, scale=1]{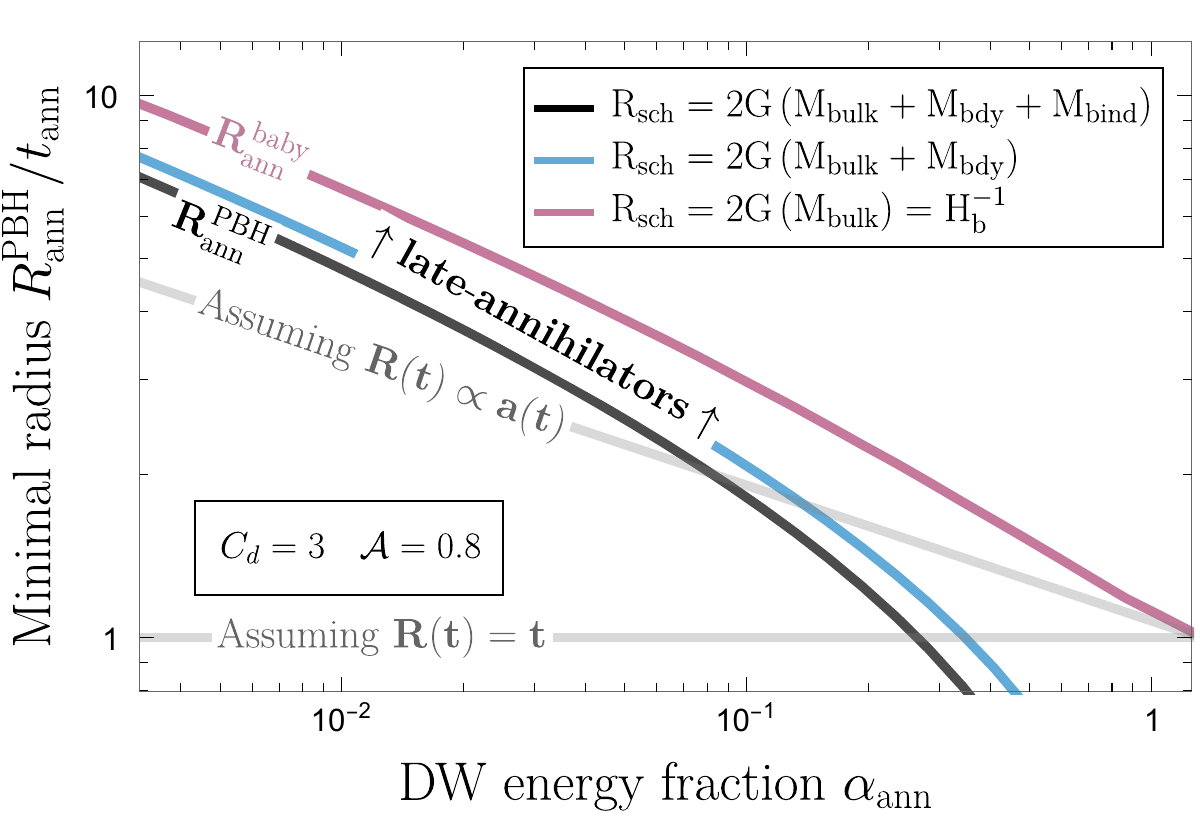}}}
\end{adjustbox}
\caption{ \small  \label{fig:Radius_EoM} {LEFT:} Time evolution of the radius $R(t)$ of a spherical DW from solving the full DW EoM for different initial sizes. DWs collapses into PBHs if they cross their Schwarzschild radius in the \textbf{blue} region. The \textbf{dark orange} line shows the trajectory of the smallest spherical DW collapsing into PBH.  The \textbf{gray} line shows the would-be DW trajectory from completely neglecting effects from the bias $V_{\rm bias}$ and surface tension $\sigma$, i.e. solely accounting for Hubble flow $R(t)\propto a(t)$. {RIGHT:}  We show the minimal radius that spherical DW must have at $t_{\rm ann}$ in order to collapse into PBH after $t_{\rm PBH}$. The Schwarzschild radius is calculated with volume term only (\textbf{purple}), plus surface term (``bdy'') (\textbf{blue}) and binding terms (\textbf{black}). Due to the relationship $2GM_{\rm bulk}=H_b^{-1}$, the \textbf{purple} line is also the critical radius to form a baby universe. The fitting formula for black and purple lines are given in Eqs.~\eqref{eq:fitting_fct_Rann} and \eqref{eq:fitting_fct_Rann_baby} respectively.}
\end{figure*}

Closed DWs collapse into PBHs at a time $t_{\rm PBH}$ after that they shrink below their Schwarzschild radius,
\begin{equation}
\label{eq:ptRadius}
 R(t_{\rm PBH})= 2G M(t_{\rm PBH}).
\end{equation}
 The mass-energy $M$ contained within the spherical DW is given by the Misner-Sharp mass~\cite{Misner:1964je,Hayward:1994bu}. The latter can be decomposed as~\cite{Deng:2020mds}
\begin{equation}
\label{eq:M_EoM}
    M=M_{\rm bulk} + M_{\rm bdy} + M_{\rm bind},
\end{equation}
where 
\begin{equation}
\label{eq:M_bulk}
    M_{\rm bulk} = \frac{4\pi}{3} V_{\rm bias}R^3,
\end{equation}
is a bulk term, 
\begin{equation}
\label{eq:M_bdy}
    M_{\rm bdy}= 4\pi \sigma \gamma R^2
\end{equation}
is a boundary term, and
\begin{equation}
     M_{\rm bind}=-8\pi^2 G\sigma^2 R^3 + 4\pi \sigma H_b\sqrt{\gamma^2-1}R^3,
\end{equation}
with $H_b = \sqrt{V_{\rm bias}/3M_{\rm Pl}^2}$, includes the repulsive surface-surface gravitational binding energy \cite{Ipser:1983db} and the attractive surface-volume gravitational binding energy, with the later typically being dominant.  
From numerically integrating the EoM in Eq.~\eqref{eq:EoM} and solving for roots of Eq.~\eqref{eq:ptRadius} assuming $R(t_{\rm PBH}) = t_{\rm PBH}$ (PBHs cannot form faster than what causality allows) we find that PBHs form after the time, see the left panel of Fig.~\ref{fig:collapse_time}:
\begin{equation}
\label{eq:t_PBH_alpha_ann}
    t_{\rm PBH} \simeq \frac{1}{2}\sqrt{\frac{3M_{\rm Pl}^2}{V_{\rm bias}}} = \frac{t_{\rm ann} }{\sqrt{C_d \alpha_{\rm ann}}}\simeq t_{\rm dom},
\end{equation}
which is around the time $t_{\rm dom}$ in Eq.~\eqref{eq:t_dom} when DW interiors become dominated by the vacuum energy $V_{\rm bias}$.
If we had approximated the DW mass by its volume term $M\simeq M_{\rm bulk}$, we would have found that PBHs are twice longer to form $t_{\rm PBH} \simeq 2t_{\rm dom}$, see the left panel of Fig.~\ref{fig:collapse_time}. Such difference can impact the PBH abundance by many orders of magnitude. In fact the bulk term composes only $25\%$ of the DW total mass at the time of collapse, see the right panel of Fig.~\ref{fig:collapse_time}.  The large surface term, composing $50\%$ of the mass budget, can be attributed to the large Lorentz factor shown in Fig.~\ref{fig:Lorentz_factor}. 

\subsection{Late-annihilators}
Typical DWs annihilate at $t_{\rm ann}$ in Eq.~\eqref{eq:t_ann} when the volume term in Eq.~\eqref{eq:M_bulk} dominates over the surface term in Eq.~\eqref{eq:M_bdy}. This argument however neglect Hubble expansion which becomes important for $R\gtrsim t$.
 In the absence of analytical solution, we solve Eqs.~\eqref{eq:EoM} numerically and display the DW trajectory in the left panel of Fig.~\ref{fig:Radius_EoM}. We find that DWs grow when their radius is super-horizon $R>t$, and quickly shrink under the work of their own surface tension and vacuum pressure when entering the horizon $R<t$.  The trajectory of the smallest spherical DWs which can collapse into PBHs is shown in dark orange. It enters its Schwarzschild radius at Hubble crossing $R\simeq t$. Larger DWs enter their Schwarzschild radius while $R>t$ (blue region) but they only start to collapse after they become causally-connected when $R\simeq t$ (black line). 
 The radius $R_{\rm ann}^{\rm PBH}$ at $t_{\rm ann}$ of the smallest DW which collapse into PBH is plotted in the right panel of Fig.~\ref{fig:Radius_EoM} as a function of $\alpha_{\rm ann}$, for different approximations for the Misner-Sharp mass in Eq.~\eqref{eq:M_EoM}. We also show in gray the approximation $R(t) \propto t^{1/2}$  only accounting for the Hubble flow and neglecting effects from the bias $V_{\rm bias}$ and surface tension $\sigma$ which become important when $R$ approaches $R\sim t$. The most precise approximation shown with the black line is well reproduced with the fitting function:
 \begin{equation}
 \label{eq:fitting_fct_Rann}
     \frac{R_{\rm ann}^{\rm PBH}}{t_{\rm ann}} \simeq 0.780\, {\rm Log_{10}^2}\left(\alpha_{\rm ann}\right) -0.618\,{\rm Log_{10}}\left(\alpha_{\rm ann}\right)  + 0.407,
 \end{equation}
 Closed DW configurations with a radius larger than $R\gtrsim R_{\rm ann}^{\rm PBH}$ at the onset of the annihilation phase at $t_{\rm ann}$ are the population of \textit{late-annihilators} which collapse into PBHs after $t_{\rm PBH}$.\footnote{Their existence was postulated in~\cite{Ferrer:2018uiu,Pujolas:2022qvs} under the name of \textit{late}-\textit{birds}. They are the DW network analogs of \textit{late}-\textit{bloomers} which are formed in the context of first-order phase transitions~\cite{Gouttenoire:2023naa,Gouttenoire:2023bqy,Gouttenoire:2023pxh}. }

\subsection{Baby-universe}
If the radius $R$ of a DW becomes larger than the cosmological horizon $H_b^{-1}$ of an inflating universe with vacuum energy density $V_{\rm bias}$, which would occur around the time $t_{\rm baby}$
\begin{equation}
\label{eq:}
    R(t_{\rm baby}) \simeq H_b^{-1},\qquad H_b\equiv \sqrt{V_{\rm bias}/3M_{\rm Pl}^2},
\end{equation}
then the DW radius grows exponentially with time, see the left panel of Fig.~\ref{fig:Radius_EoM}. This leads to the formation of an eternally inflating universe with a vacuum energy denoted by $V_{\rm bias}$ called child-universe in \cite{Sato:1981gv} and baby-universe in \cite{Giddings:1987cg,Hawking:1988fm}.

The baby-universe detaches from the parent universe through a wormhole, which closes off within a timescale comparable to its light crossing time. Observers outside the DW will continue to experience the power-law Hubble expansion characteristic of a radiation-dominated universe. They will perceive the DW as shrinking until it collapses into a PBH, see e.g. \cite{Garriga:2015fdk,Deng:2016vzb}. The observation of the DW collapsing into a PBH, while at the same time the coordinate $R(t)$ is exponentially growing, appears paradoxical. This seeming contradiction finds resolution upon recognizing that the growing DW radius $R(t)$ is in fact defined in the parallel exterior region of the maximally-extended Schwarzschild space-time diagram \cite{Blau:1986cw}. Consequently, a wormhole forms between the growing DW and the external observer, as illustrated by the space-time diagrams C, D and E in \cite{Blau:1986cw}. The surface area along the wormhole throat interpolates between the growing DW surface observed from the inside, and the surface area perceived from the outside, as depicted in Fig. 11 of \cite{Deng:2016vzb}. The same figure shows that the throat quickly pinches off, forming a BH horizon on both sides, one in the inflating baby-universe and one in the parent radiation-dominated universe. Another consequence of the wormhole geometry, demonstrated in App. C of \cite{Blau:1986cw}, is that in spite of having a growing radius $R(t)$, the DW is subject to a proper acceleration which is oriented inward, toward the expanding baby-universe.

Numerically, we find that wormholes leading to the baby-universe form roughly around twice the time of PBH formation, as given in Eq.~\eqref{eq:t_PBH_alpha_ann}:
\begin{equation}
\label{eq:t_baby}
    t_{\rm baby} \simeq 2t_{\rm PBH} \simeq 2t_{\rm dom}.
\end{equation}
Since $2GM_{\rm bulk} = H_b^2R^3$ in Eq.~\eqref{eq:M_EoM}, the time $t_{\rm baby}$ actually corresponds to the purple line in the left panel of Fig.~\ref{fig:collapse_time}.
We find that the minimal DW radius at the onset of the annihilation phase for the latter to collapse into a wormhole is well approximated by the fitting function:
\begin{equation}
 \label{eq:fitting_fct_Rann_baby}
     \frac{R_{\rm ann}^{\rm baby}}{t_{\rm ann}} \simeq 0.951\, {\rm Log_{10}^2}\left(\alpha_{\rm ann}\right) -0.860\,{\rm Log_{10}}\left(\alpha_{\rm ann}\right)  + 1.15,
 \end{equation}
which for the same reason as before corresponds to the purple line in the right panel of Fig.~\ref{fig:Radius_EoM}.  
 The abundance of late-annihilators, whose size are larger than either Eq.~\eqref{eq:fitting_fct_Rann} or Eq.~\eqref{eq:fitting_fct_Rann_baby}, depending on whether they form a simple PBH or a PBH + wormhole, is determined in the next section using percolation theory.

\section{PERCOLATION THEORY IN 3D}
\label{app:percolation_theory}
The goal of this section is to estimate the size distribution of closed DWs. In anticipation of future, appropriate numerical simulations to accurately compute this distribution, we here propose to discretize the DW network on a lattice and to use principles and results from percolation theory \cite{Stauffer:1978kr,Essam_1980,Lalak:1993bp}.
 
\subsection{Discretization}
A DW network can be viewed as a collection of domains of constant field configuration whose typical size is set by the correlation length $L$,  which in the scaling regime becomes of the order of the horizon size $L \sim t$. The correlation length is usually estimated from the network energy density \cite{Press:1989yh,Kawasaki:2014sqa,Oliveira:2004he,Martins:2016ois}:
\begin{equation}
\label{eq:correlation_length}
    \frac{L}{t} \simeq \frac{\sigma}{\rho_{\rm DW}t} = \mathcal{A}^{-1},
\end{equation}
where $\mathcal{A}$ is the area parameter introduced in Eq.~\eqref{eq:rho_DW}. Using results from field-theory lattice simulations, for $\mathcal{Z}_2$-symmetric models one finds $L/t \simeq 1.25^{+0.18}_{-0.14}$ \cite{Kawasaki:2014sqa} or $L/t \simeq 1.14\pm 0.04$ \cite{Martins:2016ois}, while $\mathcal{Z}_{N_{\rm DW}}$-symmetric models with $N_{\rm DW}>2$ are associated to a shorter correlation length, $L/t \simeq (2.70\pm 0.30)/N_{\rm DW}$ \cite{Kawasaki:2014sqa}. The values of the field $\phi(r,t)$ giving rise to the DW networks are correlated within region of size $L$ and becomes quickly uncorrelated at larger distances.
 This motivates the modeling of the DW network as a 3D periodic lattice containing $N^3$ sites with lattice spacing set equal to the correlation length $L$ \cite{Lalak:1992px,Lalak:1993bp,Lalak:1994qt,Coulson:1995nv}. 
 
 Assuming a $\mathcal{Z}_2$-symmetric model, every individual lattice sites can be in one of two states, either occupied or empty, corresponding to the field sitting in the false or true vacuum, respectively. Each site is occupied or empty entirely
randomly, independently of the state of its neighbors, with a probability $p$ or $1-p$, respectively.  In presence of an initial bias $V_{\rm bias}$, the probability of false vacuum occupation reads~\cite{Gelmini:1988sf}
\begin{equation}
\label{eq:p_Vbias}
    \frac{p}{1-p} = \exp\left(-\Delta F/T \right) \simeq \exp\left( -V_{\rm bias}/V_0 \right),
\end{equation}
where $T$ is the temperature, $\Delta F$ and $V_0$ are the free energy difference and energy density barrier between the two minima.
We assume that the potential bias is negligible at the time of DW formation $V_{\rm bias}\ll V_{0}$ such that we have $p=0.5$. The occupied sites are either isolated from one another or they form small groups of neighbors. These groups are called clusters \cite{Stauffer:1978kr,Essam_1980,Lalak:1993bp}. An $s$-cluster is defined as a group of $s$ occupied lattice sites connected by nearest neighbor distances.
Above some critical probability $p_c$ ($=0.311$ for a cubic lattice), the system undergoes a phase transition from having no infinite cluster when  $p<p_c$ to having one infinite cluster when $p>p_c$; in other words, the system becomes percolated by one continuous path of occupied sites.  This infinite cluster is unique. Apart from the infinite cluster, there will also be many finite clusters dispersed throughout the lattice. We define $n_s$ as the total number of finite clusters of size $s$ divided by the total number of lattice sites $N^3$. Hence, the total number of such clusters is  $N_s = n_s \times N^3$. The probability that any selected lattice site is part of a $s$-cluster is given by $P_s = s \times n_s$. We refer the reader to \cite{Stauffer:1978kr,Essam_1980,Lalak:1993bp} for reviews on percolation theory.

\begin{figure*}[th!]
\centering
\begin{adjustbox}{max width=1\linewidth,center}
\raisebox{0cm}{\makebox{\includegraphics[ width=0.49\textwidth, scale=1]{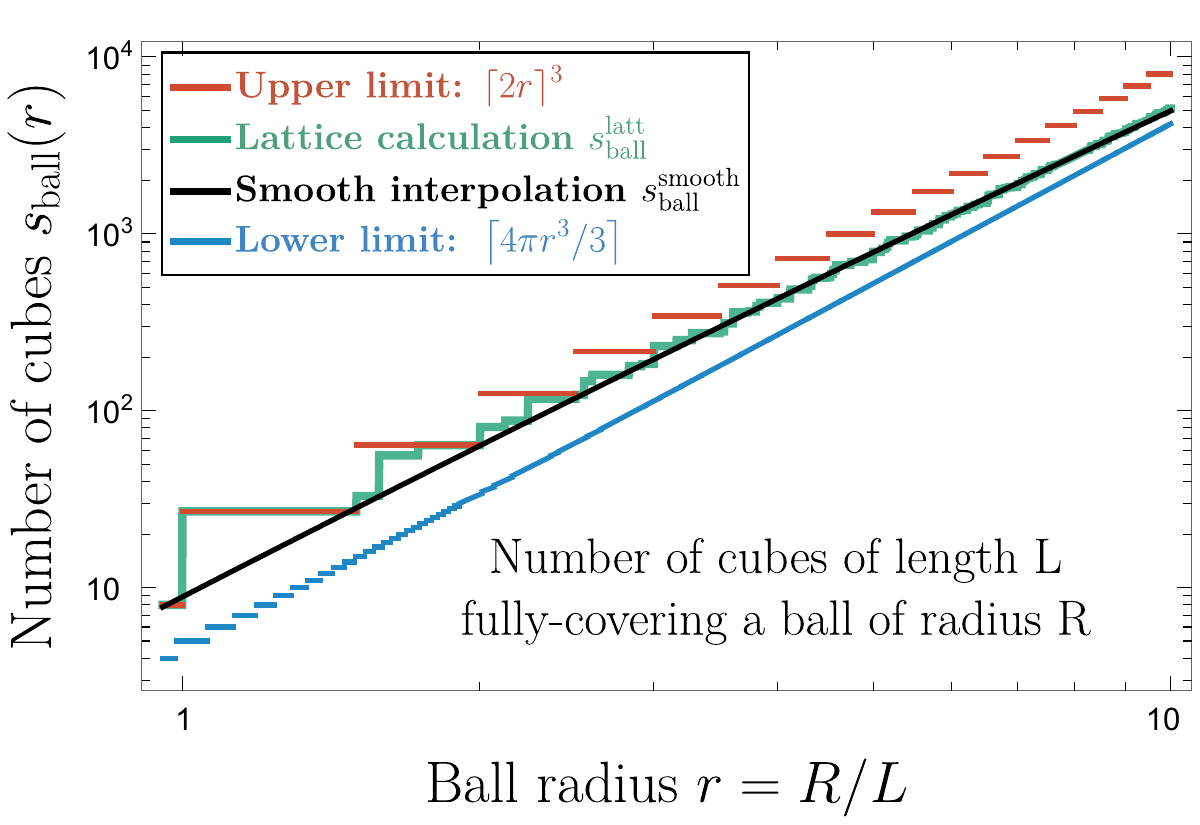}}}
\raisebox{0cm}{\makebox{\includegraphics[ width=0.49\textwidth, scale=1]{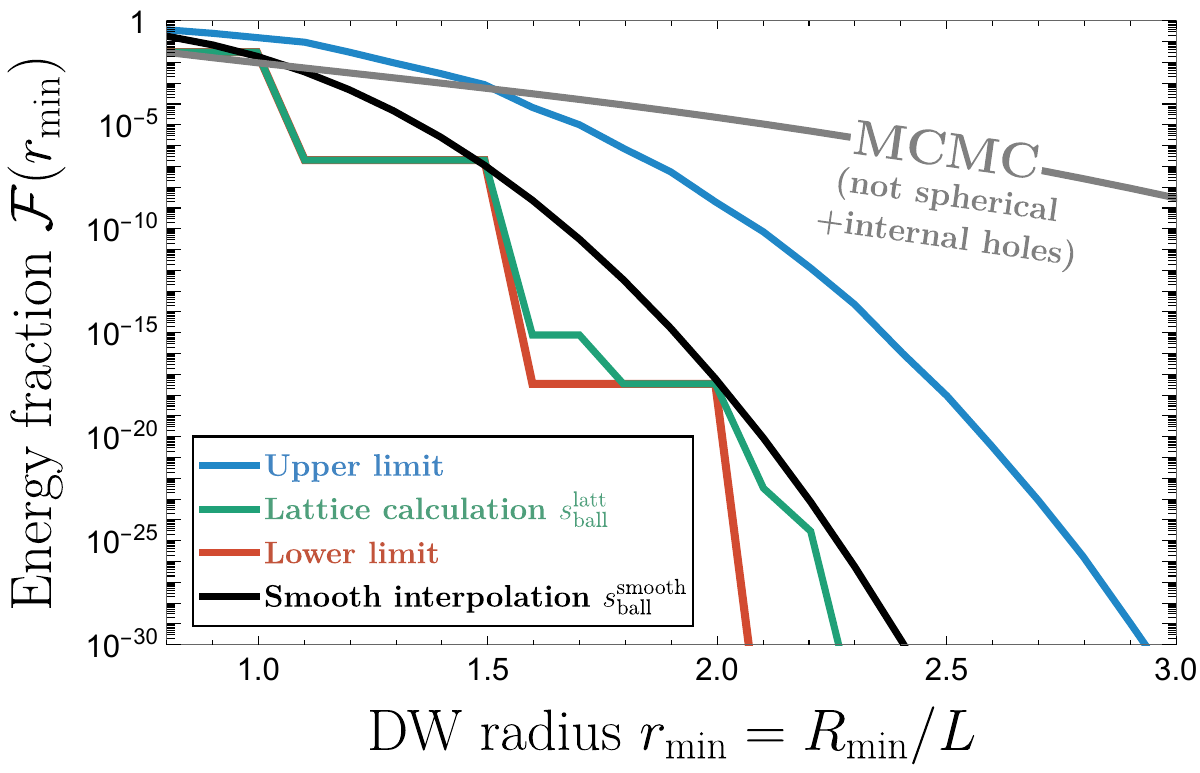}}}
\end{adjustbox}
\caption{ \small  \label{fig:F_R_vs_xi}
\textbf{Left:} Minimum number of lattice sites, denoted as $s_{\rm ball}$, required to completely cover a spherical region of radius $R$. The green line illustrates the lattice result, which is given by the analytical summations derived in App.~\ref{app:cubes_ball}. To eliminate the unphysical stair-step behavior, a smooth interpolation is applied, as depicted by the black line, with the additional condition that it must not surpass the upper limit represented by the red line, see Eq.~\eqref{eq:s_sph_spline}. As the radius $R$ increases, the number of lattice sites approaches the perfect-packing limit, indicated by the blue line. \textbf{Right:} Energy fraction of the network retained in closed DWs, which are sufficiently large to contain a false vacuum sphere with radius $R$. It is obtained by injecting the function $s_{\rm ball}(R)$ shown in the left panel in Eq.~\eqref{eq:mathcal_F_percolation_th}. The black line represents the most accurate result, achieving a smooth interpolation of the lattice data depicted in green, while consistently exceeding the lower limit, marked by the red line. In contrast, the MCMC result, presented in App.~\ref{sec:MCMC} and shown in gray, obtained from injecting Eq.~\eqref{eq:MCMC_sim} in Eq.~\eqref{eq:mathcal_F_percolation_th}, tends to overestimate the fraction $\mathcal{F}$ of DWs on the verge to collapse into PBHs as it also includes DWs  that are significantly non-spherical or that possess internal holes.  It is important to note the anti-correlation between the legends of the two panels: the lower limit in the left panel corresponds to the upper limit in the right panel, and vice versa for the upper limit. Instead, the colors match across both panels. $L$ is the network correlation length. }
\end{figure*}

\subsection{Number of late-annihilators}\label{sec:number}
In principle, the number $n_s$ of s-cluster has been calculated with Markov Chain Monte Carlo (MCMC) simulations \cite{Lalak:1993bp, 1976JPhA....9.1705S,Flammang1977,Hoshen1979,PhysRevB.22.2466} together with their typical radius $R_s$. We address this topic in App.~\ref{sec:MCMC}. However, we conclude that since these clusters tend to be irregular in shape and contain internal holes, we cannot rely on existing results from MCMC simulations to determine the fraction of DWs that collapse into PBHs. Therefore, we follow a different route which we now outline in detail. In Sec.~\ref{sec:DW_EoM}, we concluded that DWs collapsing into PBHs are the ones which are larger than a given radius $R_{\rm min}$ in Eqs.~\eqref{eq:fitting_fct_Rann} or \eqref{eq:fitting_fct_Rann_baby}. We suppose that we can relax the assumption on their spherical shape as long as they are large enough to contain a fully-occupied ball of radius $R_{\rm min}$. In doing so, any deviation from sphericity would add more mass than required for the collapse to occur.
We define $s_{\rm ball}(r_{\rm min})$ as the minimal number of lattice sites of size $L$ to fully cover a ball of radius $R_{\rm min}$, with $r_{\rm min}\equiv R_{\rm min}/L$.
The probability that all those sites are occupied is $p^{s_{\rm ball}(r_{\rm min})}$ where $p=0.5$ is the occupation probability. Multiplying it by the highest number of balls which can be packed on the full lattice $c_0N^3/s_{\rm ball}$ where $c_0=\pi/3\sqrt{2} \simeq 0.74$, we obtain the number of s-clusters large enough to contain a ball of radius $r_{\rm min}$ (in unit of $L$):
\begin{equation}
\label{eq:n_sph0}
    n_{\rm sph}(r_{\rm min})~=~d\frac{c_0}{s_{\rm ball}(r_{\rm min})} p^{s_{\rm ball}(r_{\rm min})},
\end{equation}
where we normalized by the total number of sites $N^3$.
We introduced the degeneracy factor $d$ to account for configurations that are not aligned with the arbitrarily fixed division of the lattice into $c_0N^3/s_{\text{ball}}$ sites. In the diluted limit, we anticipate that $d
\xrightarrow[s_{\text{ball}}\to \infty]{}s_{\text{ball}}/c_0$, corresponding to the number of choices for the displacement of the ball center. In the context of calculating the abundance of PBHs, we can confidently work within the diluted limit, and thus Eq.~\eqref{eq:n_sph0} becomes
\begin{equation}
\label{eq:n_sph}
    n_{\rm sph}(r_{\rm min})~\simeq~p^{s_{\rm ball}(r_{\rm min})},
\end{equation}
Those cluster carry an energy:
\begin{equation}
    E_{\rm sph} = \epsilon\times s_{\rm ball}(r_{\rm min})\times n_{\rm sph}(r_{\rm min}),
\end{equation}
where $\epsilon$ is the energy of one occupied sites. This must be compared with the energy of the DW network:
\begin{equation}
    E_{\rm DW} = \epsilon(p+p) = \epsilon,
\end{equation}
where the two terms accounts for the bulk and surface energy of the DW network which at $t_{\rm ann}$ are exactly equal. We deduce that the energy fraction contained in closed DW large enough to contain a spherical ball of radius $r_{\rm min}$ is given by:
\begin{equation}
\label{eq:mathcal_F_percolation_th}
\mathcal{F}(r_{\rm min})~=~ \frac{E_{\rm sph}}{E_{\rm DW}}~=~  s_{\rm ball}n_{\rm sph}~\simeq~ s_{\rm ball}p^{s_{\rm ball}(r_{\rm min})},
\end{equation}
where we recall $r_{\rm min}\equiv R_{\rm min}/L$ and $p=0.5$. In App.~\ref{app:cubes_ball}, we calculate the minimal number $s_{\rm ball}^{\rm latt}(r)$ of cubes of unit length required to fully cover a ball of radius $r$, under the form of analytical sums. The result is plotted with a green line in the left panel of Fig.~\ref{fig:F_R_vs_xi} against its upper and lower limit shown with red and blue lines:
\begin{equation}
\label{eq:s_sph_upper_lower_lim}
    \left\lceil\frac{4\pi r^3}{3} \right\rceil~<~ s_{\rm ball}^{\rm latt}(r) ~\leq ~\left\lceil 2r\right\rceil^3.
\end{equation}
We observe that for small radii  $r\in [1,5]$, the value adheres to its upper limit, representing a cube of length $2R$. This is due to the low number of lattice sites $\in [10, 10^3]$. Conversely, for larger radii $r \gtrsim 10$, it approaches its lower limit, representing perfect packing.
The stair-case behaviour of $s_{\rm ball}^{\rm latt}(r)$ arises because of the modeling of the field configuration on a discrete lattice. It should not reflect any physical property of the real system that should instead show a smooth behavior. This motivates the introduction of a smooth version $s_{\rm ball}^{\rm smooth}(r)$ obtained under the double condition to interpolate the analytical sum $s_{\rm ball}^{\rm latt}(r)$ in Eq.~\eqref{eq:s_ball_latt} as best as possible while keeping it always smaller than its upper limit $s_{\rm ball}^{\rm smooth}(r) ~\leq~\left\lceil 2r\right\rceil^3$.

The resulting smooth interpolation shown with black line in the left panel of Fig.~\ref{fig:F_R_vs_xi} reads:
\begin{equation}
\label{eq:s_sph_spline}
   \frac{s_{\rm ball}^{\rm smooth}(r)}{r^3} =8+\left(\frac{1+\tanh{(a_1\log{(r/a_2)})}}{2}\right)\left(\frac{4\pi}{3}-8 \right),
\end{equation}
where $a_1\simeq 1.15$, $a_2 \simeq 5.55$ and $r\equiv R/L$. The expression in Eq.~\eqref{eq:s_sph_spline} explicitly renders the asymptotics of $s_{\rm ball}^{\rm smooth}(r)/r^3$ approaching $8$ as $r \to 1$, and $4\pi/3$ as $r \to \infty$.  
The energy fraction $\mathcal{F}(r_{\rm min})$ of closed DWs larger than a ball of radius $r_{\rm min}$ is obtained from plugging $s_{\rm ball}^{\rm smooth}(r_{\rm min})$ in Eq.~\eqref{eq:s_sph_spline} into Eq.~\eqref{eq:mathcal_F_percolation_th}, and is shown with a black line in the right panel of Fig.~\ref{fig:F_R_vs_xi}. It is compared to the expression obtained from injecting the analytical sum $s_{\rm ball}^{\rm latt}(r_{\rm min})$ in Eq.~\eqref{eq:s_ball_latt} of App.~\ref{app:cubes_ball} as well as its upper and lower limits in Eq.~\eqref{eq:s_sph_upper_lower_lim}, resulting in lower and upper limits on $\mathcal{F}$ shown with red and blue lines respectively.
Additionally, we depict with a gray line the energy fraction $\mathcal{F}$ of late-annihilators which would have been obtained if results from MCMC simulations presented in Eq.~\eqref{eq:MCMC_sim} of App.~\ref{sec:MCMC} were plugged in Eq.~\eqref{eq:mathcal_F_percolation_th}.  By doing this, we would have dramatically overestimated the abundance of DW on the verge to collapse to PBH. 

\section{PBH AND WORMHOLE ABUNDANCE}
\label{sec:PBH_WH_abundance}

Now that we are in possession of the critical size $R_{\rm min}^{\rm PBH
}$ and $R_{\rm min}^{\rm baby
}$ beyond which closed DW collapse into PBH and wormholes, derived in Sec.~\ref{sec:DW_EoM}, and of the DW size distribution derived in Sec.~\ref{app:percolation_theory}, we can now proceed to the derivation of the PBH and wormhole abundance.

\subsection{PBHs}
 The PBH contribution to the DM abundance is
 \begin{equation}
 \label{eq:f_PBH_def}
    f_{\rm PBH} = \frac{\rho_{\rm coll}(t_{\rm PBH})}{\rho_{\rm DM}(t_{\rm PBH})}=\frac{\rho_{\rm DW}(t_{\rm ann})}{\rho_{\rm DM}(t_{\rm PBH})}\frac{\rho_{\rm coll}(t_{\rm PBH})}{\rho_{\rm coll}(t_{\rm ann})}\mathcal{F}_{\rm coll},
\end{equation}
where $\rho_{\rm DM}(t)$ is the DM energy density, $\rho_{\rm coll}(t)$ is the energy density stored in collapsing DWs and $\mathcal{F}_{\rm coll}$ is the collapsing fraction at the onset of annihilation,
\begin{equation}
   \mathcal{F}_{\rm coll} \equiv \frac{\rho_{\rm coll}(t_{\rm ann})}{\rho_{\rm DW}(t_{\rm ann})}.
\end{equation}
The first factor in right side of Eq.~\eqref{eq:f_PBH_def} can be evaluated from evolving DWs, DM and radiation energy densities like $t^{-1}$, $a^{-3}$ and $g_\star(T)T^4$ respectively and matching them at $t_{\rm dom}^{\mathsmaller{\rm unbias}}$ in Eq.~\eqref{eq:t_dom_no_bias} and matter-radiation equality when $T_{\rm eq}\simeq 0.80~\rm eV$.
The second factor in Eq.~\eqref{eq:f_PBH_def} can be evaluated from using that DWs collapsing into PBHs have super-horizon size and therefore their energy evolves approximately as $a^{-1}$. One obtains:
\begin{equation}
\label{eq:f_PBH_final}
    f_{\rm PBH} ~=~
  \mathcal{G} \times \mathcal{R} \times \frac{T_{\rm dom}}{T_{\rm eq}}\times \mathcal{F}_{\rm coll},
\end{equation}
with
\begin{equation}
\label{eq:mathcal_G}
     \mathcal{G}\equiv \left(\frac{g_{s\star}^{\rm eq}}{g_\star^{\rm eq}} \right)\left(\frac{g_{s\star}^{\rm PBH}}{g_{s\star}^{\rm ann}} \right)^{\!\!1/3}\hspace{-0.2cm}
   \left( \frac{(g_\star^{\rm ann}g^{\rm dom}_\star)^{\rm 1/4}}{g_{s\star}^{\rm PBH}/(g^{\rm PBH}_\star)^{\! 1/2}}\right) \sim 1,
\end{equation}
and 
\begin{equation}
\label{eq:mathcal_R}
\mathcal{R} \simeq ~ \left(\frac{t_{\rm PBH}}{t_{\rm dom}^{\mathsmaller{\rm unbias}}} \right) \left(\frac{t_{\rm dom}}{t_{\rm ann}} \right)^{\! 1/2} \sim 1. 
\end{equation}
The different temperatures can be calculated from the characteristic times in Eqs.\,(\ref{eq:t_ann},\ref{eq:t_dom},\ref{eq:t_dom_no_bias},\ref{eq:ptRadius}) using $T\simeq 1.23(M_{\rm Pl}/t)^{1/2}/g_\star(T)^{1/4}$, with $g_\star(T)$ and $g_{*s}(T)$ the number of relativistic degrees of freedom appearing in energy and entropy density respectively, written as $g_{*}^X$ and $g_{s,*}^X$ with $X$ the associated epoch in Eq.~\eqref{eq:mathcal_G}. In Eqs.~\eqref{eq:mathcal_G} and \eqref{eq:mathcal_R}, behind the sign $\sim$ we anticipated that PBH formation reaches maximum efficiency when DWs have the longest lifespan. This happens when they annihilate just before dominating the universe $t_{\rm ann}\sim t_{\rm dom} \sim t_{\rm dom}^{\mathsmaller{\rm unbias}}$, which together with Eq.~\eqref{eq:t_PBH_alpha_ann}, implies $\mathcal{G}\sim 1$ and $\mathcal{R}\sim 1$.
The collapsing fraction $\mathcal{F}_{\rm coll}$ is given by the fraction of closed DWs in Eq.~\eqref{eq:mathcal_F_percolation_th} with a radius larger than the critical threshold:
\begin{equation}
\label{eq:mathcal_F_percolation_th_2}
\mathcal{F}_{\rm coll}=\mathcal{F}\left(r_{\rm ann}^{\rm PBH}\right)~\simeq~  s_{\rm ball}\left(r_{\rm ann}^{\rm PBH}\right)\times p^{s_{\rm ball}\left(r_{\rm ann}^{\rm PBH}\right)},
\end{equation}
where $p=0.5$ and $r_{\rm ann}^{\rm PBH}\equiv R_{\rm ann}^{\rm PBH}/L$. The function $s_{\rm ball}(r=R/L)$, given in Eq.~\eqref{eq:s_sph_spline}, is the number of correlated regions within a ball of radius $R$. The network correlation length $L$ must be evaluated at $t_{\rm ann}$ just before the annihilation stage starts, thus in the scaling regime. Possible values for $L$ are discussed below Eq.~\eqref{eq:correlation_length}. The critical radius $R_{\rm ann}^{\rm PBH}$ at $t_{\rm ann}$ beyond which DWs are expected to collapse into PBHs is given by the fitting function in Eq.~\eqref{eq:fitting_fct_Rann}. 


For rapid use, we propose the following fitting function for Eq.~\eqref{eq:mathcal_F_percolation_th_2}:
\begin{equation}
\label{eq:F_coll_fit}
    \mathcal{F}_{\rm coll}\simeq \exp\left[- \frac{a}{\ell^b}\left(\frac{1}{\alpha_{\rm ann}}\right)^{\dfrac{c}{\ell^{d}}}\right],
\end{equation}
with $a\simeq 0.659$, $b\simeq 2.49$, $c\simeq 1.61$, and $d=0.195$. We checked that Eq.~\eqref{eq:F_coll_fit} provides a very good approximation of Eq.~\eqref{eq:mathcal_F_percolation_th_2}, after substitution of Eqs.~\eqref{eq:fitting_fct_Rann} and \eqref{eq:s_sph_spline}, for $\ell \equiv L/t$ within the range $[0.2,2]$.
The PBH abundance in shown with brown lines in  Fig.~\ref{fig:DW_PBH_constraints} and Fig.~\ref{fig:DW_PBH_constraints_GW_xi}. 
The minimal PBH mass is given by the mass inside the Schwarzschild radius at horizon crossing $R_{\rm sch}(t_{\rm PBH})=t_{\rm PBH}$: 
\begin{equation} \label{eq:MPBH}
    M_{\rm PBH} \simeq \frac{t_{\rm PBH}}{2G}\simeq M_{\mathTerra} \left( \frac{217~\rm GeV}{V_{\rm bias}^{1/4}} \right)^{2},
\end{equation}
where $t_{\rm PBH}$ is given by Eq.~\eqref{eq:t_PBH_alpha_ann} and $M_{\mathTerra} \simeq 3.00\times 10^{-6}~M_{\mathSun}$ where $M_{\mathTerra}$ and $M_{\mathSun}$ are Earth and Sun masses.   The PBH mass can also be calculated as $M_{\rm PBH}\simeq 4\times 4\pi t_{\rm PBH}^3V_{\rm bias}/3$, which is the bulk mass inside the horizon multiplied by a factor $4$ accounting for the surface and gravitational binding energies, see the right panel of Fig.~\ref{fig:collapse_time}. In principle, PBH produced above the threshold $R\gg R_{\rm ann}^{\rm PBH}$ have larger mass than the minimal value in Eq.~\eqref{eq:MPBH}. The production of heavier PBH is exponentially suppressed by the late-annihilator fraction in Eq.~\eqref{eq:mathcal_F_percolation_th}, thus we expect a nearly-monochromatic mass distribution for PBHs produced by DW networks. We leave its precise calculation for future works. In Fig.~\ref{fig:DW_PBH_constraints}, we recast the usual PBHs astrophysical constraints in the parameter space of DW networks.

\subsection{Wormholes}
The critical DW radius $R_{\rm min}^{\rm  baby}$ in Eq.~\eqref{eq:fitting_fct_Rann_baby} beyond which the collapse into PBHs is also associated to the creation of a wormhole connected to an inflating baby-universe is about $30\%$ times larger than the critical radius $R_{\rm min}^{\rm  PBH}$ in Eq.~\eqref{eq:fitting_fct_Rann} for forming simple PBHs. Due to the exponential suppression in Eq.~\eqref{eq:mathcal_F_percolation_th}, this implies that the wormhole production rate is significantly lower than the PBH production rate. In spite of this suppression, is it possible that at least one baby universe has been created within our past light cone? 
The number $f_{\rm baby}$ of baby universes formed in our past lightcone is simply:
\begin{equation}
f_{\rm baby} =\mathcal{N}_{\rm patches}(T_{\rm dom}) \times \mathcal{F}(R_{\rm ann}^{\rm baby}),
\end{equation}
where the DW fraction $\mathcal{F}(R_{\rm min})$ is defined in Eqs.~\eqref{eq:mathcal_F_percolation_th} and \eqref{eq:s_sph_spline}, and the baby threshold $R_{\rm ann}^{\rm baby}$ is given in Eq.~\eqref{eq:fitting_fct_Rann_baby}. The function $\mathcal{N}_{\rm patches}(T_{\rm dom})$ is the number of Hubble patches, at the time of collapse in Eqs.~\eqref{eq:t_baby} and \eqref{eq:t_dom} when the temperature is approximately $T_{\rm dom}$, in our past light-cone. It reads:
\begin{align}
\label{eq:N_patches}
\mathcal{N}_{\rm patches}  
&= \left( \frac{a_{\rm dom} H_{\rm dom}}{a_0 H_{0}} \right)^3\notag \\
&\simeq \frac{1.3 \times 10^{38}}{h^3} \left(\frac{g_\star(T_{\rm dom})}{100}\right)^{1/2}\left( \frac{T_{\rm dom}}{100~\rm GeV} \right)^{3},
\end{align}
where $h\equiv H_0/100~\rm km/s/Mpc$ with $H_0$ the Hubble constant today. We approximated $g_{*}(T)\simeq g_{*,s}(T)$ and used $g_{*s}(T_{0})\simeq 3.94$. The DW network parameter space producing at least one wormhole connected to baby universe is indicated with the dashed purple line in Fig.~\ref{fig:DW_PBH_constraints} and Fig.~\ref{fig:DW_PBH_constraints_GW_xi}.

 {
\begin{figure*}
\centering
\begin{adjustbox}{max width=1\linewidth,center}
\raisebox{0cm}{\makebox{\includegraphics[ width=0.9\textwidth, scale=1]{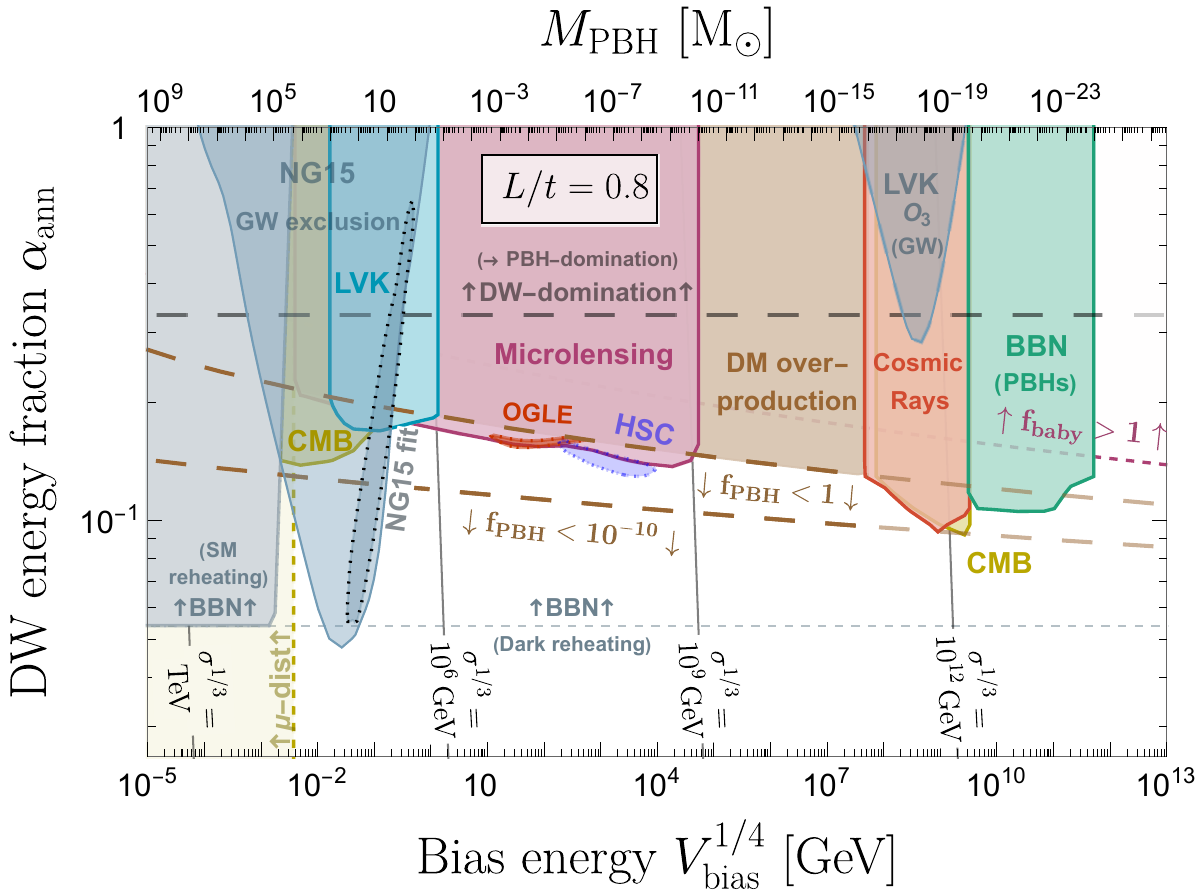}}}
\end{adjustbox}
\caption{ \small  \label{fig:DW_PBH_constraints} Exclusion constraints on DW network occupying an energy fraction $\alpha_{\rm ann}$ defined in Eq.~\eqref{eq:alpha_ann} at the onset of the annihilation phase driven by a bias energy density difference $V_{\rm bias}$ between distinct vacua. PBHs are produced around the temperature $V_{\rm bias}^{1/4}$ shown on the bottom x-axis with the PBH mass shown on the top x-axis. The {\bf colorful} regions are excluded due to the presence of PBHs.
The two {\bf yellow} areas on the {\bf left} side indicate regions excluded by the Cosmic Microwave Background (CMB) either due to $\mu$-distortion caused by inhomogeneities accompanying the production of very large PBH shown in dashed~\cite{Kohri:2014lza,Nakama:2016kfq,Nakama:2017xvq}, or due to the accretion of large PBHs shown in solid~\cite{Ali-Haimoud:2016mbv,Poulin:2017bwe,Serpico:2020ehh}.
The region in {\bf cyan} is ruled out due to limits on the black hole merging rate from LIGO-Virgo-Kagra (LVK)~\cite{DeLuca:2020qqa}. The {\bf purple} region is excluded by the constraints from MACHO~\cite{MACHO:2000qbb}, Eros~\cite{EROS-2:2006ryy}, OGLE~\cite{Niikura:2019kqi}, and HSC~\cite{Niikura:2017zjd} microlensing experiments. The red and purple dotted {\bf horizontal ellipses} show the best-fit regions that can explain the anomalies observed in OGLE and HSC microlensing data \cite{Niikura:2017zjd,Niikura:2019kqi,Sugiyama:2021xqg}.
The {\bf brown} area shows where DM would over-close the universe. Above the {\bf purple} dashed line, more than one baby-universe $f_{\rm baby}\gtrsim 1$ is produced in our past lightcone. We only show it in the region not excluded by PBH overproduction. At smaller masses, PBH evaporate within a universe lifetime \cite{Hawking:1974rv,Hawking:1975vcx}.
Since we do not observe the presence of Hawking radiation, neither in terms of cosmic-ray fluxes~\cite{Carr:2009jm,Boudaud:2018hqb,Laha:2019ssq,DeRocco:2019fjq,Laha:2020ivk}, nor in terms of modification of the ionisation fraction in CMB~\cite{Poulin:2016anj,Stocker:2018avm,Poulter:2019ooo}, nor in terms of modificaton of the abundance of light elements produced during Big-Bang Nucleosynthesis (BBN)~\cite{Carr:2009jm}, we can exclude the {\bf red}, {\bf yellow} and {\bf green} regions on the right side respectively. The {\bf gray} region labelled ``DW domination'' defined by $t_{\rm ann}\gtrsim t_{\rm dom}$, see Eq.~\eqref{eq:alpha_ann}, indicates where the bias vacuum energy becomes larger than the radiation density of the universe. Such region is expected to lead to a PBH-dominated universe as mentioned in the introduction. Additionally, the {\bf blue-gray} regions indicate where stochastic GW background (SGWB) produced from DW annihilation are excluded by NANOGrav 15-year (NG15) data \cite{NANOGrav:2023ctt} and $O_3$-run of LIGO-Virgo-Kagra (LVK)~\cite{KAGRA:2021kbb}. We use that no SGWB with a signal-to-noise ratio larger than $\rm SNR\leq 5$ has been detected in NG15 and $\rm SNR\leq 2$ for LVK. The blue-filled black dotted {\bf vertical ellipse} shows the $90\%$ favoured region which can explain the SGWB recently detected in NG15 \cite{NANOGrav:2023gor,Antoniadis:2023rey,Reardon:2023gzh,Xu:2023wog,InternationalPulsarTimingArray:2023mzf}, using the Bayesian analysis of \cite{Gouttenoire:2023ftk}.
The regions labeled ``BBN'' in {\bf gray}, are excluded by the constraints on number of effective degrees of freedom $N_{\rm eff}\lesssim 0.4$ \cite{Pitrou:2018cgg,Dvorkin:2022jyg}. Two scenarios must be distinguished according to whether the DW network annihilates into degrees of freedom from the Standard Model (SM) or from a dark sector, see App.~\ref{app:BBN} for the details. Lastly, the DW network correlation length has been set to $L =0.8 t$, which is a rather conservative assumption for $\mathcal{Z}_2$-symmetric network, see Eq.~\eqref{eq:correlation_length}.}
\end{figure*}
\clearpage
}

\subsection{Further considerations}
\label{sec:further_consideration}
Finally, we discuss additional effects which could impact the PBHs abundance predicted in this work, as well as aspects reserved for future investigation.

\textbf{Collapse of sub-horizon PBHs.}
In this study, we only account for PBHs formed from DWs entering their Schwarzschild radius at the moment of their horizon entry or before.
We do not account for the possibility for DWs to enter their Schwarzschild radius much after their horizon entry, at a radius
$R_{\rm sch}/t \sim (t/2t_{\rm dom})^2$, possibly much smaller than $t$, where we assumed $R_{\rm sch}=2GM$ with $M\sim 4\pi t^3/3$.
 PBH formation well within the horizon can only occur if they are spherical enough to shrink by a factor $(2t_{\rm dom}/t)^2\gg 1$ to become fully contained within their Schwarzschild sphere. The PBH abundance from sub-horizon collapse is highly sensitive to the distribution of DW shapes. For this reason, in this analysis we made the conservative choice to not account for this additional contribution, which we leave for future research.
 
\textbf{Deviation from spherical symmetry.}
In Sec.~\ref{sec:DW_EoM}, we calculate the critical radius $R_{\rm ann}^{\rm PBH}$ beyond which DWs collapse into PBH, see e.g. Eq.~\eqref{eq:M_EoM}. The fact that DWs are in general not spherical raises the question whether this treatment is appropriate to describe the real physical system.
However, we apply this criteria on closed DW which are large enough to contain a ball of radius $R_{\rm min}$, their abundance $\mathcal{F}(R_{\rm min})$ being derived in Sec.~\ref{app:percolation_theory}. 
If $R_{\rm min}=R_{\rm ann}^{\rm PBH}$, those DWs will all unavoidably collapse into PBHs whatever their shape. This is because non-sphericity will participate to a mass larger than $M(R_{\rm min})$ in Eq.~\eqref{eq:M_EoM} and therefore make it easier to collapse into PBHs. Instead, our analysis is in fact conservative with respect to non-sphericity since it misses DW configurations which are not accounted by $\mathcal{F}(R_{\rm min})$ due to their non-spherical shapes in spite of being successful at forming PBHs.

\textbf{Internal holes.} Our analysis focuses on calculating the amount of completely false-vacuum-dominated spherical configurations. Consequently, it excludes false vacuum configurations that could collapse into PBHs despite possessing internal cavities of the true vacuum phase. The study of PBH formation via cheese-like configurations is postponed to future research, with the anticipation that such corrections are likely to increase the amount of PBHs.

\textbf{Correlation length.}
The PBH abundance shows an exponential sensitivity to the correlation length $L$ of the DW network, as indicated in Eq.~\eqref{eq:F_coll_fit}. In this study, we suggest estimating $L$ through the DW network energy density, with $L \approx \sigma/\rho_{\rm DW}$, as presented in Eq.~\eqref{eq:correlation_length}. This approach benefits from previous lattice calculations, where the relationship $L/t \approx (2.70\pm 0.30)/N_{\rm DW}$ was established for the $\mathcal{Z}_{N_{\rm DW}}$-symmetric model \cite{Kawasaki:2014sqa}. This result implies that the PBH abundance dramatically decreases as the number $N_{\rm DW}$ of nearly-degenerate vacua increases. A more precise determination of the correlation length, along with its dependency on $N_{\rm DW}$, is left for future studies. Also the precise dependence of the late-annihilator fraction as a function of the number of vacua is left for future studies. The question at hand is whether the fraction of late annihilators experiences a significant reduction for $N_{\text{DW}} > 2$ due to the substitution of $p = 0.5$ with $p = 1/N_{\text{DW}}$ in Eq.~\eqref{eq:mathcal_F_percolation_th}.

\textbf{Friction.} The assumption of a significant source of friction -- which is considered negligible in this study but may become relevant in certain scenarios \cite{Blasi:2022ayo,Blasi:2023sej} -- would slow down the motion of DW, thereby reducing the correlation length $L$, and preventing it from keeping pace with the cosmic horizon $L \sim t$~\cite{Martins:1995tg}. This could subsequently lower the resulting PBH abundance, see e.g. the right panel of Fig.~\ref{fig:F_R_vs_xi}. Nevertheless, DWs that are larger than the cosmic horizon should remain unaffected by friction, suggesting that PBH formation should still occur in the presence of friction. In any case, whether or not friction is active, as mentioned in the introduction if the DW network were to dominate the energy density of the universe, one should anticipate a substantial production of PBHs, including those that host a baby universe.

\textbf{Presence of an initial bias.}
 Our analysis assumes that $V_{\rm bias}$ is absent at the time of network formation. The presence of a large initial bias $V_{\rm bias}$ already active at the time of network formation~\cite{Coulson:1995nv,Hindmarsh:1996xv,Pujolas:2022qvs}, i.e. such that $t_{\rm ann}<t_{\rm form}$, would impact the occupation probability $p$ in Eq.~\eqref{eq:p_Vbias} and reduce the fraction $\mathcal{F}$ of late-annihilators in Eq.~\eqref{eq:mathcal_F_percolation_th_2} and the resulting PBH abundance. Also models of DW with time-decreasing surface tension~\cite{Babichev:2021uvl,Ramazanov:2021eya,Babichev:2023pbf} would give different results.

\begin{figure*}[th]
\centering
\begin{adjustbox}{max width=1\linewidth,center}
\raisebox{0cm}{\makebox{\includegraphics[ width=1\textwidth, scale=1]{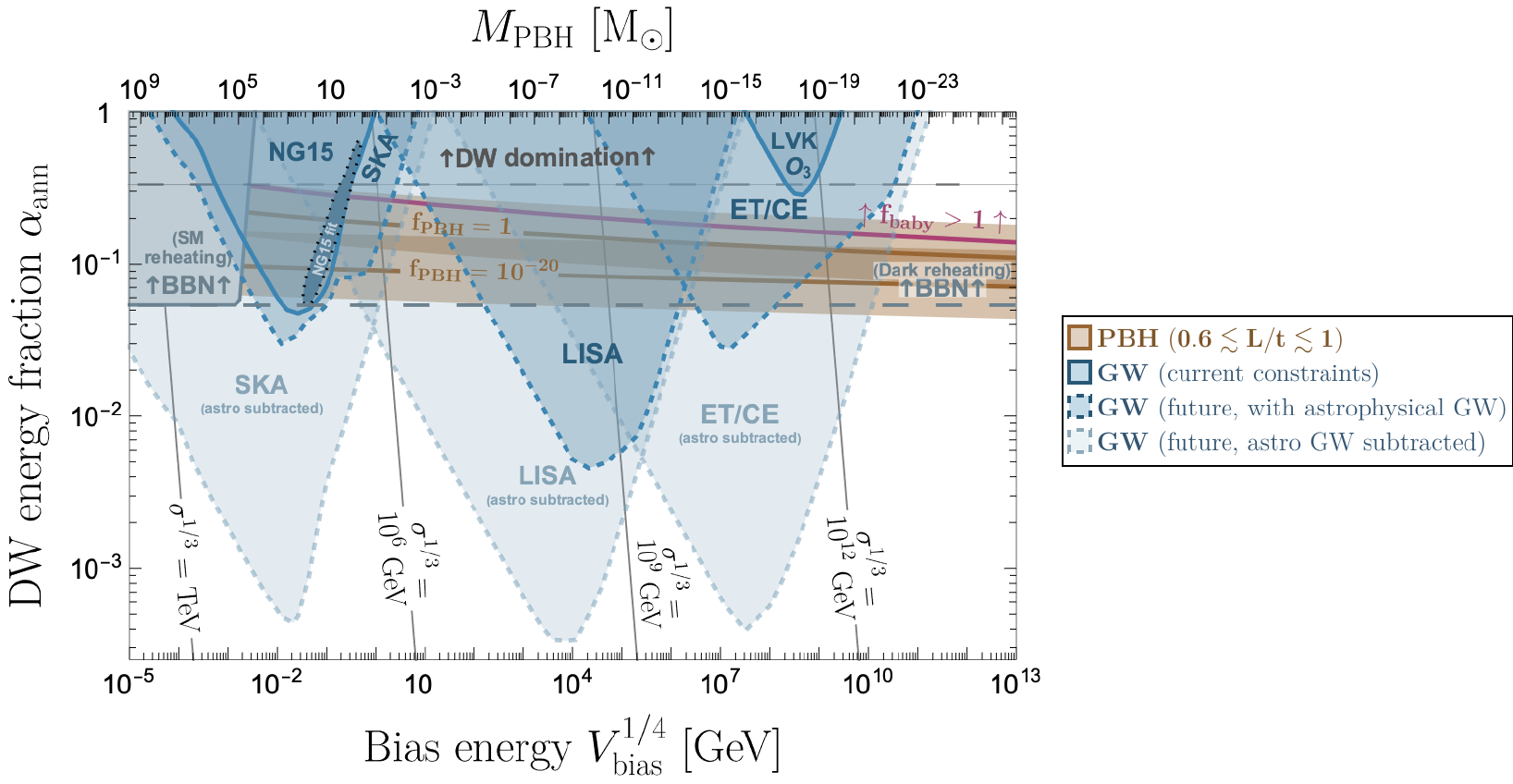}}}
\end{adjustbox}
\caption{ \small  \label{fig:DW_PBH_constraints_GW_xi} Observability of stochastic GW background (SGWB) produced by annihilating DW networks (\textbf{blue}) compared to PBH abundance (\textbf{brown}). 
 Solid blue lines indicate existing GW constraints from NANOGrav 15-year data release (NG15) \cite{NANOGrav:2023ctt} and run $O_3$ from LIGO-Virgo-Kagra (LVK) \cite{KAGRA:2021kbb}, assuming the Signal-to-Noise Ratio detection thresholds $\rm SNR =5$ and $\rm SNR =2$, respectively. The dashed blue lines indicate the future prospects from SKA \cite{Janssen:2014dka}, LISA \cite{Audley:2017drz,Robson:2018ifk,LISACosmologyWorkingGroup:2022jok} and ET/CE \cite{Punturo:2010zz,Maggiore:2019uih,Reitze:2019iox}. The low opacity regions show the most optimistic prospects, assuming a low detection threshold $\rm SNR =1$ (equivalent to $1\sigma$ deviation from noise) after $T=20~\rm years$ of observation ($T=10~\rm years$ for SKA) \cite{Schmitz:2020syl} and the perfect subtraction of all the stochastic astrophysical GW foregrounds, see e.g. Fig.~6 in \cite{Baldes:2023fsp}. Instead the higher opacity regions show the more conservative prospects assuming no astrophysical foreground subtraction.
The thickness of the brown bands represents the uncertainty on the correlation length $L$ of the DW network $L/t \in [0.6,1]$, the solid line showing the central value $L=0.8t$. Above the \textbf{purple} line, at least one eternally inflating baby-universe is produced in our past light cone, assuming $L=0.8t$. The regions labelled ``BBN'' and ``DW domination'' are the same as in Fig.~\ref{fig:DW_PBH_constraints}.}
\end{figure*}

\section{GRAVITATIONAL WAVES}
\label{sec:GW}
During the annihilation process, DWs are driven to relativistic speed and radiate GWs (see e.g.~\cite{Vilenkin:1981zs,Preskill:1991kd,Chang:1998tb,Gleiser:1998na,Hiramatsu:2012sc,Gelmini:2021yzu,Li:2023gil}). 
The GW power spectrum today $\Omega_{\rm GW}^0$ can be related to the GW power spectrum at the annihilation epoch $\Omega_{\rm GW}^{\rm ann}$ by:
\begin{align}
\label{eq:_Omega_GW_0_DW}
\Omega_{\rm GW}^0h^2 = \mathcal{D}\, \Omega_{\rm GW}^{\rm ann},
\end{align}
where $\mathcal{D}\equiv \rho_{\rm rad}(T_{\rm ann})/\rho_{\rm rad}(T_{0})$ is the redshift factor of the universe radiation energy density $\rho_{\rm rad}(T)=\pi^2 g_*T^4/30$ assuming an adiabatic evolution $T\propto g_{s,*}^{-1/3}a^{-1}$:
\begin{align}
    \mathcal{D} &= \Omega_{\rm rad}^{0}h^2\left( \frac{g_{*}(T_{\rm ann})}{g_\star(T_0)}\right)\left( \frac{g_{s,*}(T_0)}{g_{s,*}(T_{\rm ann})}\right)^{4/3}\notag\\
    &\simeq 1.62 \times 10^{-5}\left( \frac{g_{*}(T_{\rm ann})}{106.75}\right)\left( \frac{106.75}{g_{s,*}(T_{\rm ann})}\right)^{4/3},
\end{align}
 We introduced today's radiation energy fraction $\Omega_{\rm rad}^{0}h^2=h^2\rho_{\rm rad}(T_0)/\rho_0 \simeq 4.18\times 10^{-5}$ with $T_0 \simeq 2.73~\rm K$ \cite{ParticleDataGroup:2022pth}, today's critical density $\rho_0=3M_{\rm Pl}^2H_0^2$ with $H_0\simeq 100h~\rm km/s/Mpc$, $g_{*}(T_0)\simeq 3.38$ and $g_{s,\star}(T_0)\simeq 3.94$, assuming $N_{\rm eff}\simeq 3.045$ \cite{Mangano:2005cc, deSalas:2016ztq}.
The GW spectrum produced by long-lived DWs annihilating at $t_{\rm ann}$ follows from the quadrupole formula~\cite{Hiramatsu:2010yz,Kawasaki:2011vv,Hiramatsu:2013qaa,Saikawa:2017hiv}:
\begin{equation}
\Omega_{\rm GW}^{\rm ann} = \epsilon_{\mathsmaller{\rm GW}} \frac{G\mathcal{A}^2\sigma^2}{\rho_{\rm rad}(t_{\rm ann})} S(f) = \frac{3}{32\pi}\epsilon_{\mathsmaller{\rm GW}}\alpha_{\rm ann}^2 S(f).
\end{equation}
The dimensionless quantity $\epsilon_{\mathsmaller{\rm GW}}$ encodes the deviation from the quadrupole formula and is fitted on lattice simulations $\epsilon_{\mathsmaller{\rm GW}}\simeq 0.7 \pm 0.4$~\cite{Hiramatsu:2013qaa}.
The spectral function $S(f)$ must have an IR slope $\Omega_{\rm GW}\propto f^{3}$ to respect causality~\cite{Durrer:2003ja,Caprini:2009fx,Cai:2019cdl,Hook:2020phx} and a UV slope $\Omega_{\rm GW}\propto f^{-1}$ as suggested by lattice simulations results~\cite{Hiramatsu:2013qaa} (though more complicated spectra are possible, see e.g.~\cite{Gelmini:2020bqg}). This motivates the modelling of $S(f)$ by the following smoothing function
\begin{equation}
S(f) = \frac{2}{\left(f/f_{\rm peak}\right)+\left(f_{\rm peak}/f\right)^3}.
\end{equation}
The peak frequency is given the Hubble factor at the time of annihilation redshifted until today \cite{Hiramatsu:2013qaa}:
\begin{multline}
f_{\rm peak} = \frac{a(t_{\rm ann})}{a(t_0)} H(t_{\rm ann}) \simeq 1.6~{\rm mHz}\left( \frac{g_\star(T_{\rm ann})}{106.75} \right)^{1/2}\\
\times\left( \frac{106.75}{g_{*s}(T_{\rm ann})}\right)^{1/3}\left(\frac{T_{\rm ann}}{10~\rm TeV} \right).
\end{multline} 
In Fig.~\ref{fig:DW_PBH_constraints}, we show the exclusion regions due to the limit on Stochastic GW Background (SGWB) by Pulsar Timing Arrays NANOGrav \cite{NANOGrav:2023ctt}, and earth-based interometers LIGO-Virgo-Kagra (run O3) \cite{KAGRA:2021kbb}. 
In Fig.~\ref{fig:DW_PBH_constraints_GW_xi}, we also present the potential for detection by future pulsar timing array SKA \cite{Janssen:2014dka}, as well as forthcoming space-based interferometers LISA \cite{Audley:2017drz,Robson:2018ifk}, ET \cite{Punturo:2010zz,Maggiore:2019uih}, and CE \cite{Reitze:2019iox}. We have set the Signal-to-Noise Ratios at $\rm SNR = 5$ and $\rm SNR = 2$ for NG15 and LVK O3, respectively, and $\rm(SNR=1,~ T=10 ~years)$ for SKA, as well as $\rm (SNR=1, ~T=20~ years)$ for LISA, ET, and CE, where $T$ is the observation time. The power-law integrated curves (PLIC) are adopted from \cite{Schmitz:2020syl}. For CE and ET, we choose the minimum of the two PLIC.  For NG15, we show both the posterior -- labelled ``NG15 fit'' -- assuming that the recently detected GW signal \cite{NANOGrav:2023gor,Antoniadis:2023rey,Reardon:2023gzh,Xu:2023wog,InternationalPulsarTimingArray:2023mzf} is sourced by annihilating DWs \cite{NANOGrav:2023hvm,Gouttenoire:2023ftk}, and the exclusion constraints -- labelled ``GW exclusion'' -- which we would have had if no GW would have been detected \cite{NANOGrav:2023ctt}. We can interpret the latter as the current NG15 constraints on annihilating DWs assuming that the origin of the GW signal is imputed to supermassive black hole binaries. 
For the future GW prospects, we explore two different scenarios, depending on whether the expected astrophysical foreground could be subtracted or not. We take into account the galactic white dwarf binaries, referencing either the model from \cite{Lamberts:2019nyk,Boileau:2021gbr} or the one from \cite{Robson:2018ifk}, as well as extragalactic supermassive black hole binaries from \cite{Rosado:2011kv}, and extragalactic compact binaries (comprising of neutron stars and black holes) fitted on LIGO O3 data \cite{KAGRA:2021kbb}. We refer to Fig.~6 in \cite{Baldes:2023fsp} for a visualisation of the different GW astrophysical foregrounds across frequencies.  The shaded regions in Figs.~\ref{fig:DW_PBH_constraints} and \ref{fig:DW_PBH_constraints_GW_xi} correspond to places where the GW background from DW in Eq.~\eqref{eq:_Omega_GW_0_DW} exceeds the corresponding PLIC plus the eventual GW astrophysical foregrounds.

\section{CONCLUSION}

Domain Wall (DW) networks could have formed in the early universe after the spontaneous breaking of a $\mathcal{Z}_N$ symmetry with $N>1$. In presence of a vacuum energy difference $V_{\rm bias}$ lifting the degeneracy between the $N$ vacua, DWs are driven toward annihilating each other. In order to be viable, the DW network must annihilate before occupying a significant energy fraction of the universe. The Schwarzschild radius of DWs grows with radius as $R_{\rm sch}\propto R^2$ or $R^3$ according to whether surface or bulk energy dominates. The average size of DWs, which is set by the correlation length $L$ of the network, grows with cosmic time as $L\simeq t$.  This seems to imply that there must be a time $t_{\rm PBH}$ when most of the DWs enter their Schwarzschild radius and collapse into PBHs. This time is closely related the time $t_{\rm dom}\sim t_{\rm PBH}$ when the universe becomes dominated by the DW network energy density. This is the recipe followed by previous papers \cite{Ferrer:2018uiu,Gelmini:2022nim,Gelmini:2023ngs} to calculate the PBH abundance from DW networks. 
However, as first pointed in \cite{Gelmini:2023kvo}, the DW growth cannot be applied during the annihilation phase which necessarily precedes PBH formation $t_{\rm ann} \lesssim t_{\rm PBH}$ to prevent DW domination.  In this work, we find that only super-horizon DWs continue to grow like $R\propto a(t)$ or slower while sub-horizon DWs shrink. This implies that only super-horizon DWs which are larger than a given threshold, $R\geq R_{\rm ann}^{\rm PBH}$, can enter the horizon after $t_{\rm PBH}$ and collapse into PBHs. We call this population \textit{late-annihilators}. We calculate the PBH formation threshold $R_{\rm ann}^{\rm PBH}$ accounting for volume, surface, and gravitational binding energies arising from Einstein equations. We introduce a new formalism borrowed from percolation theory in 3D to calculate the abundance of late-annihilators which we find to be exponentially suppressed for small DW energy fraction $\alpha_{\rm ann}$ or small network correlation length $L$, see Eq.~\eqref{eq:F_coll_fit}.

We are able to translate PBH exclusion constraints to the parameter space of DW networks in Fig.~\ref{fig:DW_PBH_constraints}.
We find that the newly-constrained parameter space overlaps with the regions which are either currently probed by GW observatories or will be probed in the future, see Fig.~\ref{fig:DW_PBH_constraints_GW_xi}.
We conclude that Bayesian analysis of DW network interpretation of stochastic GW background in current and future observatories should account for constraints from PBH production as initiated in our companion paper \cite{Gouttenoire:2023ftk}.

Another novelty of the present study is to have determined the conditions for DWs to continue expanding forever as seen from an observer located inside. Those eternally inflating baby-universes are connected to their parent universe through a wormhole. The latter pinches off in one light crossing time, leaving a PBH behind, with no residual evidence for the baby-universe genesis. Those eternally-inflating baby universes are in principle allowed to undergo nucleation of bubbles of any other vacua permitted by the underlying particle physics \cite{Garriga:2015fdk,Garriga:1997ef,Linde:2015edk}. Hence, baby-universes will evolve into complex space-time structures, comprising a multitude of eternally-inflating regions interconnected by wormholes, giving rise to a Multiverse. We find that the possibility to have a baby-universe formed in our past lightcone is excluded by PBH over-production, except in the region where PBH evaporate before the onset of BBN if they are lighter than $10^{-24}~M_{\mathSun}\sim 10^9~\rm g$. This occurs if the bias energy density is larger than $V_{\rm bias} \gtrsim (10^{12}~\rm GeV)^4$, see Figs.~\ref{fig:DW_PBH_constraints} and \ref{fig:DW_PBH_constraints_GW_xi}.

Finally, the $\mu$-distortion constraints shown in dashed yellow region in Fig.~\ref{fig:DW_PBH_constraints} assume that inhomogeneities sourced by the DW networks are not highly non-Gaussian \cite{Nakama:2017xvq} in contrast to the claim in \cite{Ferrer:2018uiu}. Those  $\mu$-distortion constraints could jeopardize the possibility for PBH to be produced from QCD axion cosmological scenarios as initially proposed in \cite{Ferrer:2018uiu}. We leave more quantitative conclusions for future studies.

\section*{Acknowledgments}
The authors thank the anonymous referees for the thorough review.
YG thanks Aleksandr Azatov, Marek Lewicki, Ken'ichi Saikawa, Peera Simakachorn, and Ville Vaskonen for helpful discussions, and is grateful to the Azrieli
Foundation for the award of an Azrieli Fellowship.
EV acknowledges support by
the European Research Council (ERC) under the European Union’s Horizon Europe research and innovation
programme (grant agreement No. 101040019).

\appendix



\begin{figure*}[th]
\centering
\begin{adjustbox}{max width=1\linewidth,center}
\raisebox{0cm}{\makebox{\includegraphics[ width=0.8\textwidth, scale=1]{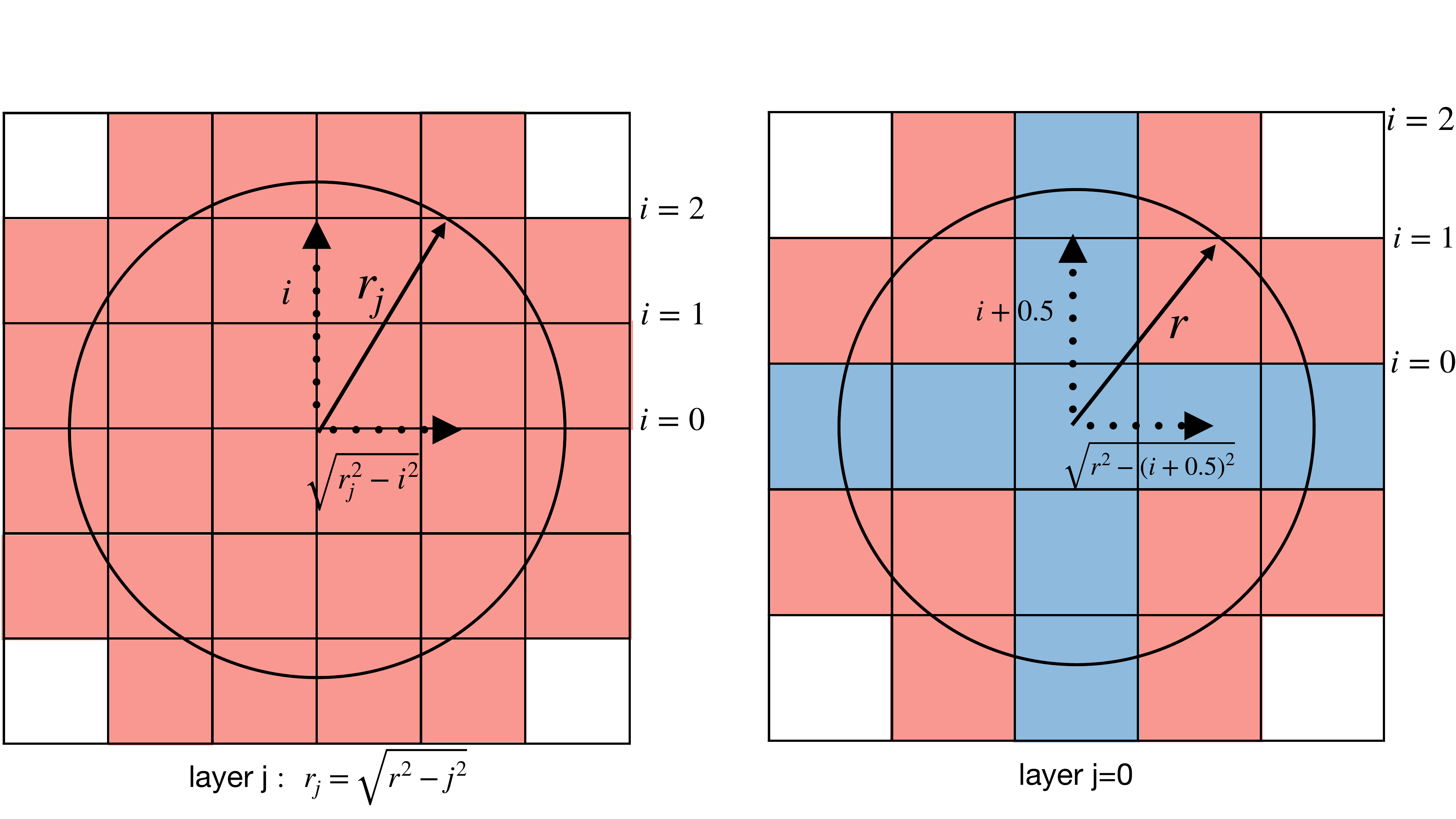}}}
\end{adjustbox}
\caption{ \small  \label{fig:cubes_ball} We consider two possible way to pave a ball with cubes, according to whether the ball center is at the center or a corner of a cube, whose 2D projections are shown in the \textbf{left} and \textbf{right} panel respectively.. The corresponding numbers of cubes fully covering the ball are denoted $s_{\rm ball,1}(r)$ and $s_{\rm ball,2}(r)$.  }
\end{figure*}
\section{MCMC simulations}
\label{sec:MCMC}
The number of $s$-cluster has been calculated with Markov Chain Monte Carlo (MCMC) simulations in both condensed matter~\cite{1976JPhA....9.1705S,Flammang1977,Hoshen1979,PhysRevB.22.2466} and cosmological context~\cite{Lalak:1993bp}. One finds~\cite{Lalak:1993bp}
\begin{equation}
\label{eq:n_s}
    n_s = \frac{0.0501}{s^\tau}\exp\left\{-0.6299\mathcal{P}s^\sigma(\mathcal{P}s^\sigma + 1.6679) \right\},
\end{equation}
where $\tau \simeq 2.17$, $\sigma \simeq 0.48$, $\mathcal{P}\equiv (p-p_c)/p_c$. For $p>p_c$, the  average cluster radius $R_s$ is given by \cite{Lalak:1993bp}
\begin{equation}
\label{eq:R_s}
    R_s/L  = 0.702|p-p_c|^{-0.124} s^{1/3}.
\end{equation}
For an occupation probability $p=0.5$ and the percolation threshold of a cubic lattice $p_c=0.311$, Eqs.~\eqref{eq:n_s} and \eqref{eq:R_s} simplify to
\begin{equation}
\label{eq:MCMC_sim}
    n_s \simeq 0.0501 s^{-2.17}\exp{-0.2326s^{0.96}-0.6385s^{0.48}},
\end{equation}
and $R_s = 0.863 s^{1/3}$. Since $p>p_c$, most of the lattice points belong to the infinite cluster. We note $P_\infty$ the probability that an occupied site belongs to this percolating cluster. A given lattice point can either be empty with a probability $1-p$, occupied in the percolating cluster with a probability $pP_\infty$, or occupied in a finite cluster with a probability $\sum_s sn_s $. Summing all the possibilities leads to \cite{Stauffer:1978kr}
\begin{equation}
\label{eq:sum_proba}
    1- p + pP_\infty +\sum_s sn_s = 1.
\end{equation}
Plugging Eq.~\eqref{eq:MCMC_sim} into Eq.~\eqref{eq:sum_proba} for $p=0.5$, we calculate $P_\infty \simeq 0.92$. For the cosmological system ($N\to +\infty$), it means that $92\%$ of the false vacuum region (with vacuum energy $V_{\rm bias}$) is contained inside an infinite DW spanning the whole universe and only $8\%$ of the false vacuum region is contained inside finite, closed, DWs.
The number of these closed DWs is represented by $n_s$ in Eq.~\eqref{eq:MCMC_sim}. However, the formula given in Eq.~\eqref{eq:MCMC_sim} is not suitable for estimating the number of DWs collapsing into PBHs due to two considerations.

Firstly, the analysis conducted in Sec.~\ref{sec:DW_EoM} presumes that the DWs are spherical. The scaling relationship between cluster volume and radius $s \propto R_s^3$ in Eq.~\eqref{eq:R_s} suggests that clusters exhibit a droplet-like shape, as elaborated in \cite{Stauffer:1978kr}. Nonetheless, when observing the simulated shapes of DWs on a lattice, e.g. as in Fig.~1 of \cite{Press:1989yh}, it becomes evident that typical closed DWs can deviate significantly from a spherical shape at larger sizes. This deviation can be characterized by their inner radius being potentially much smaller than their average radius $R_s$.
Secondly, at larger sizes, clusters tend to encompass a considerable number of internal holes, similar to Swiss cheese \cite{Stauffer:1978kr}. In light of these factors, we deduce that the number distribution $n_s$, derived from MCMC simulations in Eq.~\eqref{eq:MCMC_sim}, which accounts for clusters that are significantly irregular in shape and contain internal holes, would overestimate the number of DWs on the verge of collapse.  
Statistical methods to determine the shape of an s-cluster, for instance the classical Wulff construction in the supercritical regime where the surface energy is minimized for a given volume (e.g. \cite{Alexander:1989sj,dobrushin1992wulff,cerf2006wulff}), are beyond the scope of this study.
In this work, we determine the fraction of DWs collapsing into PBHs with our own methods as indicated in Sec.~\ref{sec:number}. The outcomes displayed in the right panel of Fig.~\ref{fig:F_R_vs_xi} indeed indicate that results from MCMC simulations found in the literature would significantly overestimate the fraction of DWs undergoing collapse.\footnote{ Note that Fig.~\ref{fig:F_R_vs_xi}-right displays Eq.~\eqref{eq:n_s} for $p=0.5$ and $s\lesssim 42$, derived from substituting $R_s\lesssim 3$ into Eq.~\eqref{eq:R_s}. This corresponds to $\mathcal{P}s^{\sigma} \lesssim 3.66$, potentially exceeding the valid application range of Eq.~\eqref{eq:n_s}. Upon reading Ref.~\cite{Lalak:1993bp}, it remains ambiguous whether the validity regime of Eq.~\eqref{eq:n_s} is $\mathcal{P}s^{\sigma} \lesssim 5.79$ or $\mathcal{P}s^{\sigma} \lesssim 1.41$. Regardless, the conclusion that the results from MCMC simulations like Eq.~\eqref{eq:n_s} are unsuitable for estimating the PBHs abundance remains the same.}

\section{Number of cubes to cover a ball}
\label{app:cubes_ball}
In this appendix, we calculate the minimal number of cubes of length $L$ required to fully cover a ball of radius $R$, or equivalently the number of cubes of unit length covering a ball of radius $r=R/L$. We propose the two configurations shown in Fig.~\ref{fig:cubes_ball} where the center of the ball is either a face or a center of a cube.
We start with the latter. We can divide the ball into  $8$ octants, which we subsequently divide into $\lceil r\rceil$ layers denoted by $j\in \Iintv{0,\lfloor r \rfloor}$, one of them being shown in the left panel of Fig.~\ref{fig:cubes_ball}. The radius of the circle coinciding with the boundary of the ball at a layer $j$ is $r_j=\sqrt{r^2-j^2}$. Denoting the column number by $i\in \Iintv{0,\lfloor r \rfloor}$, the number of cubes filling a row, shown in red color, is $\left\lceil \sqrt{r_j^2 -i^2}\right\rceil$. We deduce the total number of cubes filling the ball for the configuration shown in the left panel of Fig.~\ref{fig:cubes_ball}:
\begin{equation}
    s_{\rm ball,1}(r)= 8\sum_{j=0}^{\lfloor r \rfloor}\sum_{i=0}^{\lfloor \sqrt{r^2-j^2} \rfloor} \left\lceil \sqrt{r^2 -i^2-j^2}\right\rceil.
\end{equation}
We now pass to the next configuration where the ball center coincides with the center of a cube. We start by isolating the 3 orthogonal layers along the planes $(XY)$, $(XZ)$ and $(YZ)$ passing through the center of the ball, one of them being shown in the right panel of Fig.~\ref{fig:cubes_ball}. The number of cubes in the central tri-axis blue cross is $6\lfloor r+0.5 \rfloor +1$. The remaining number of cubes in those three orthogonal layer outside the 3-axis cross is $4\sum_{i=0}^{\lfloor r \rfloor} \left\lceil \sqrt{r^2 -(i+0.5)^2}+0.5\right\rceil$. 
Finally, accounting for the remaining cubes in the 8 octants outside the three internal layers give:
\begin{widetext}
\begin{equation}
     s_{\rm ball,2}(r)=6\lfloor r+0.5 \rfloor +1+ 4\sum_{i=0}^{\lfloor r \rfloor} \left\lceil \sqrt{r^2 -(i+0.5)^2}+0.5\right\rceil+ 8\sum_{j=0}^{\lfloor r \rfloor}\sum_{i=0}^{\lfloor \sqrt{r^2-j^2} \rfloor} \left\lceil \sqrt{r^2 -(i+0.5)^2-(j+0.5)^2}\right\rceil.
\end{equation}
\end{widetext}
The minimal number of cubes to fully cover a ball obtained from the analytical sums of this appendix is
\begin{equation}
\label{eq:s_ball_latt}
    s_{\rm ball}^{\rm latt}(r)=\textrm{Min}\left[s_{\rm ball,1}(r),s_{\rm ball,2}(r)\right].
\end{equation}
The function $s_{\rm ball}^{\rm latt}(r)$ with $r=R/L$ is shown with a green line in Fig.~\ref{fig:F_R_vs_xi}. We find that at small radius $\gtrsim 1$, it is close to its upper limit shown in red for which the ball is replaced by its outer-cube of length $2r$
\begin{equation}
    s_{\rm ball}^{\rm latt}(r) \gtrsim \left\lceil 2r \right\rceil^3.
\end{equation}
At large radius $r\gg 1$, the number of cubes asymptotes the perfect-packing (lower) limit shown in blue for which the tiny cubes smoothly fit inside the ball:
\begin{equation}
      s_{\rm ball}^{\rm latt}(r) \lesssim \left\lceil \frac{4\pi r^3}{3} \right\rceil.
\end{equation}

\section{BBN bound} 
\label{app:BBN}
Domain walls (DWs) form a component of the total energy density of the universe. As such, they contribute to increase the expansion rate of the universe which makes neutron freeze-out earlier, increase the $n/p$ ratio which in turn increases the Helium abundance \cite{Baumann:2022mni}. The presence of DWs can be described in terms of an extra number of neutrino species
\begin{equation}
\label{eq:Delta_Neff}
 N_{\rm eff} = \frac{8}{7}\left( \frac{\rho_{\rm DW}}{\rho_\gamma} \right)\left( \frac{11}{4} \right)^{4/3},
\end{equation}
where $\rho_{\gamma}$ is the photon number density.  We introduce the DW energy fraction in unit of radiation energy density at annihilation temperature
\begin{equation}
\label{eq:alpha_tot_app}
    \alpha_{\rm DW}(T) = \frac{\rho_{\rm DW}(T)}{\frac{\pi^2}{30}g_\star(T)T^4}.
\end{equation}
where $T$ is the SM photon temperature.
From Eq.~\eqref{eq:Delta_Neff} and Eq.~\eqref{eq:alpha_tot_app}, the maximal DW contribution to $N_{\rm eff}$ occurs at annihilation temperature 
\begin{equation}
\label{eq:Delta_Neff_bound}
    \Delta N_{\rm eff}(T) =  \frac{8}{7}\left( \frac{{g_{*}(T)}}{2} \right)\left( \frac{11}{4} \right)^{4/3}\!\!\left(\frac{g_{*,s}(T_0)}{g_{*,s}(T)} \right)^{4/3}\!\!\alpha_{\rm DW}(T),
\end{equation}
To apply the BBN bound $\Delta N_{\rm eff}  \lesssim 0.3$ \cite{Pitrou:2018cgg,Dvorkin:2022jyg}, the effective number of extra relativistic degrees of freedom must be evaluated below the neutrino decoupling temperature where $g_\star(T< T_{\rm dec})\equiv 2+(7/8)\cdot 2\cdot N_{\rm eff}\cdot (4/11)^{4/3} \simeq 3.38$ and $g_{\star,s}(T< T_{\rm dec})\equiv 2+(7/8)\cdot 2\cdot N_{\rm eff}\cdot (4/11) \simeq 3.94$ where we used $N_{\rm eff}\simeq 3.045$ \cite{Mangano:2005cc, deSalas:2016ztq}. Hence, we obtain
\begin{equation}
\label{eq:Delta_N_eff_app}
    \Delta N_{\rm eff} = 7.4~\left( \frac{{g_{*}(T)}}{g_{*}(T_0)} \right)\left(\frac{g_{*,s}(T_0)}{g_{*,s}(T)} \right)^{4/3}\!\!\alpha_{\rm GW}~\lesssim~0.4.
\end{equation}
As discussed in the main text, we must distinguish the scenario in which DWs reheat to dark radiation, in which case Eq.~\eqref{eq:Delta_N_eff_app} is the BBN constraints, from the scenario in which DWs reheat to SM, in which case Eq.~\eqref{eq:Delta_N_eff_app} applies only if DWs annihilate below the neutrino decoupling temperature $T_{\rm ann} \lesssim 1~\rm MeV$.

\bibliography{biblio}
\end{document}